\pdfoutput=1
\RequirePackage{fix-cm}
\documentclass{svjour3}                     
\smartqed
\usepackage{graphicx,subfig,epsfig,color}
\usepackage{hyperref,lineno,latexsym,courier,multirow,upgreek,commath}
\hypersetup{colorlinks,citecolor=red,linkcolor=blue,urlcolor=red}

\journalname{Transport in Porous Media}

\begin{document}
\title{Effective Rheology of Two-phase Flow in Three-Dimensional Porous Media: Experiment and Simulation}

\titlerunning{Two-phase Flow in 3D Porous Media}  
\author{
  Santanu Sinha
  \and
  Andrew T. Bender
  \and
  Matthew Danczyk
  \and
  Kayla Keepseagle
  \and
  Cody A. Prather
  \and
  Joshua M. Bray
  \and
  Linn W. Thrane
  \and
  Joseph D Seymour
  \and
  Sarah L Codd
  \and
  Alex Hansen
}

\institute{
  Santanu Sinha \at
  Beijing Computational Science Research Center,
  10 East Xibeiwang Road, Haidian District, Beijing 100193, China.\\
  \email{santanu@csrc.ac.cn}\\
  \and
  Alex Hansen \at
  Beijing Computational Science Research Center,
  10 East Xibeiwang Road, Haidian District, Beijing 100193, China.\\
  Department of Physics, Norwegian University of Science and Technology, NTNU,
  N-7491 Trondheim, Norway.\\
  \email{alex.hansen@ntnu.no}\\
  \and
  Andrew T. Bender
  \and
  Matthew Danczyk
  \and
  Kayla Keepseagle
  \and
  Cody A. Prather
  \and
  Joshua M. Bray
  \and
  Linn W. Thrane
  \and
  Sarah L Codd \at
  Department of Mechanical and Industrial Engineering,
  Montana State University, Bozeman, MT, USA.\\
  \email{scodd@montana.edu}
  \and
  Joseph D Seymour \at
  Department of Chemical and Biological Engineering,
  Montana State University, Bozeman, MT, USA.\\
  \email{jseymour@montana.edu}
}

\date{Received: date / Accepted: date}

\maketitle


\begin{abstract}
  We present an experimental and numerical study of immiscible
  two-phase flow in 3-dimensional (3D) porous media to find the
  relationship between the volumetric flow rate ($Q$) and the total
  pressure difference ($\Delta P$) in the steady state. We show that
  in the regime where capillary forces compete with the viscous
  forces, the distribution of capillary barriers at the interfaces
  effectively creates a yield threshold, making the fluids reminiscent
  of a Bingham viscoplastic fluid in the porous medium, introducing a
  threshold pressure $P_t$. In this regime, $Q$ depends quadratically
  on an excess pressure drop ($\Delta P-P_t$). While increasing the
  flow-rate, there is a transition, beyond which the flow is Newtonian
  and the relationship is linear. In our experiments, we build a model
  porous medium using a column of glass beads transporting two fluids
  -- de-ionized water and air. For the numerical study, reconstructed
  3D pore-networks from real core samples are considered and the
  transport of wetting and non-wetting fluids through the network are
  modeled by tracking the fluid interfaces with time. We find
  agreement between our numerical and experimental results. Our
  results match the mean-field results reported earlier.


\end{abstract}


\section{Introduction}
\label{sec:intro}
The simultaneous flow of two immiscible fluids in porous media,
otherwise known as two-phase flow \cite{bear72,dullien92}, is getting
increasing attention of both the scientific and industrial
communities. These flows are encountered in many industrial and
geophysical applications, such as carbon sequestration and oil
recovery, groundwater management, blood flow in capillary vessels,
catalyst supports used in automotive industry, bubble generation in
microfluidics and many more. Extensive study of hydrodynamics in
porous media is crucial for the development of these applied
processes. While single-phase flows have been well characterized, many
phenomena of steady-state immiscible two-phase flows in porous media
are still not adequately understood.

In single-phase flow, the macro-scale pressure gradient ($\Delta P$)
over a porous medium scales linearly with the superficial fluid
velocity governed by Darcy's law \cite{darcy56,whitaker86}. This is
also true in the case of two-phase flows at high fluid velocities when
capillary forces are negligible. However, at slow velocities when the
capillary forces are comparable with the viscous forces, the total
pressure gradient in the steady state does not scale linearly with the
flow rate. In a fundamental experiment of two-phase flow of air and
water-glycerol mixture in a two-dimensional (2D) model porous media
made of glass beads in Hele-Shaw cell
\cite{tallakstad09,tallakstad09b}, it was observed that $\Delta P$
scales with the flow rate by a power law with an exponent $0.54$. In a
different experiment of two-phase flow in a three-dimensional (3D)
porous medium, similar power law scaling between $Q$ and $\Delta P$
was observed, but the exponents were found to vary in the range
between $0.45$ to $0.3$ depending on the saturation \cite{rassi11}.

This question about the non-linear relationship between $\Delta P$ and
$Q$ in steady-state two-phase flow of immiscible Newtonian fluid was
addressed in detail by Sinha {\it et al.}
\cite{sinhaepl12,sinhapre13}. It was shown that the capillary barriers
at the interfaces between the fluids effectively create an effective
yield threshold, making the fluids reminiscent of a Bingham
viscoplastic fluid in the porous medium introducing an overall
threshold pressure $P_t$ in the system. A generalized Darcy equation
was then obtained with mean-field calculations, where $Q$ depends
quadratically on an excess pressure drop ($\Delta P-P_t$) for low
capillary numbers ($\text{Ca}$). The capillary number is a
dimensionless number which describes different regimes of flow,
defined as the ratio of the viscous to the capillary forces at the
pore level given by $\text{Ca}=\mu v/\gamma$. Here $\mu$ is the
effective viscosity of the system, $\gamma$ is the surface tension and
$v$ is the fluid velocity which is equal to the flow rate per
cross-sectional pore area. When other parameters are constant,
$\text{Ca}$ will therefore be proportional to the flow rate, $Q$.

For a single capillary tube with a narrow pore-throat, there will be
no flow through it as long as the pressure drop ($\Delta p$) across
the tube is less than the capillary barrier ($p_t$). Above the
threshold pressure, the average flow rate $\langle q\rangle$ through
one tube can be obtained for the steady-state conditions by
integrating the instantaneous linear two-phase flow equation over the
whole tube. The problem becomes equivalent to a forced over-damped
oscillator, providing a threshold pressure followed by a square-root
singularity given by \cite{sinhapre13},
\begin{equation}
  \label{singletubeeq}
  \displaystyle
  \left<q\right> = - \sigma_0\ {\rm sgn}(\Delta p)
  \left\{  \begin{array}{cl} 
    \sqrt{\Delta p^2-p_t^2} & \mbox{if $|\Delta p| >   p_t$}\;,\\
    0                             & \mbox{if $|\Delta p| \le p_t$}\;,\\
  \end{array}
  \right.
\end{equation}
where ${\rm sgn}()$ is the sign function and $\sigma_0$ contains the
terms related to permeability. Close to the threshold, when $|\Delta
p| \ll p_t$, the average flow $\langle q\rangle\sim\sqrt{|\Delta
  p|-p_t}$. This relationship leads to the non-linear conductivity of
a single capillary given by $\sigma(\Delta p)=-\dif q/\dif (\Delta p)$
in the steady-state.\footnote{Steady state signifies that the
  macroscopic parameters describing the flow remain constant. It does
  {\it not\/} imply that the interfaces in the pores do not move,
  split or merge.}  Using this expression of $\sigma(\Delta p)$ for
each link, a mean-field theory was developed to find the relationship
of the total flow rate for a homogeneous disordered network
\cite{sinhaepl12}. For a spatially uncorrelated distribution of
threshold pressures, one obtains
\begin{equation}
  \label{effdarcy}
  Q = -C\frac{A}{L}\frac{K(S_\text{nw})}{\mu(S_\text{nw})}{\rm sgn}\left(\Delta P\right)\left\{
  \begin{array}{cl}
    \left(|\Delta P|-P_t(S_\text{nw})\right)^2 & \mbox{if $|\Delta P| >   P_t$}\;,\\
    0                                        & \mbox{if $|\Delta P| \le P_t$}\;,\\
\end{array}
\right.
\end{equation}
where $L$ is length of the network, $C$ is a constant with units of
inverse pressure and $K(S_\text{nw})$ is the effective permeability as
a function of the non-wetting fluid saturation $S_\text{nw}$. $\mu$ is
the effective viscosity of the system given by
$\mu=\mu_\text{nw}S_\text{nw}+\mu_\text{w}(1-S_\text{nw})$ where
$\mu_\text{w}$ and $\mu_\text{nw}$ are the viscosities of the wetting
and the non-wetting fluids respectively. For the high flow rates,
there is a transition, beyond which the two-phase flow rate varies
linearly with the total pressure drop. This quadratic relationship
between $Q$ and the excess pressure drop ($\Delta P-P_t$) at low
$\text{Ca}$ and the crossover to the Newtonian flow regime at high
$\text{Ca}$ were also verified with extensive numerical
simulations. However, the simulations were performed only for regular
2D networks of disordered tubes and not for 3D pore networks.

This situation of two-phase flow of two immiscible Newtonian fluids in
a disordered network is similar to the flow of a Bingham viscoplastic
fluid in a disorder network \cite{roux87,talon15}. For a Bingham fluid
inside a uniform tube, no fluid flows as long as the pressure across
the tube is less than a threshold value and the flow is linear above
the threshold. There are more complicated non-trivial flow regimes
that exist for Bingham fluid flowing through a rough channel depending
on the roughness \cite{talon14}. When a Bingham fluid flows through a
network of uniform tubes with a distribution of disorder in the
thresholds, it was shown that there is an overall threshold pressure
($P_t$) for the whole network. The threshold pressure along one
continuous flow-path throughout the entire system is the sum of all
the thresholds along that path and the minimum sum along all such
possible paths corresponds to the threshold pressure $P_t$ for the
entire system. The fluid starts flowing beyond $P_t$ and any further
increase in the pressure difference across the network by a value
$\dif P$ will cause an increase in the number of conducting links
$\dif N$ in the network. With a reasonably smooth distribution of
thresholds, we can consider $\dif N\propto\dif P$ and the change in
the conductance $\Sigma$ of the network can then be written as
$\dif\Sigma\propto\dif N\propto\dif P$. The increase in the flow rate
$\dif Q\propto \Sigma\dif P$, and integration over these will lead to
the same quadratic relationship shown in equation \ref{effdarcy}
\cite{roux87}. When all the links are conducting beyond a certain
pressure drop, $\Sigma$ becomes constant and then there is the
crossover where $Q$ becomes linear with $P$. Notice that, the
flow-rate through one link above the threshold pressure varies
linearly with the excess pressure drop for a Bingham fluid whereas in
two-phase flow through a single varying-diameter capillary it is
non-linear, but still the two phase flow in a disorder network
effectively behaves similar to a Bingham viscoplastic fluid in the
steady state.

The experiments on two-phase flow
\cite{tallakstad09,tallakstad09b,rassi11} have shown this non-linear
quadratic dependence between flow rate and pressure drop but the
existence of the threshold pressure was not investigated in any
experiment. In support of the experimental results observed in the 2D
experiment, a scaling theory was proposed in \cite{tallakstad09b}
showing $Q\sim\Delta P^2$ without considering a threshold pressure
drop. The scaling theory was based on the assumption that the fluids
flow through different channels and there are stuck clusters in
between them. The scaling theory is as follows. Let us consider a 2D
porous media of width $W$ and length $L$ where the overall pressure
drop as well as the total flow is in the direction of length
$L$. There is a number of flow channels $n_l$ in the direction of
overall flow. A characteristic length-scale $l$ in between the
channels is considered and therefore $n_l = W/l$. The flow through
each channel is then, $q_c = ka^2\Delta P/(\mu L)$, where $a$ is the
cross section of one channel and $k$ is the permeability of a channel.
The total flow through all the channels is then $Q = n_lq_c =
ka^2W\Delta P/(\mu lL)$. In order to find the dependence of $l$ on
$\Delta P$, it is assumed that the stuck clusters in between the
channels are held in place by capillary forces $p_c$. If $\lambda$ is
the diameter of such a cluster, then the viscous pressure drop around
it will be $\lambda\Delta P/L$, which should be $\le p_c$ in order for
the cluster to be held in place. Considering $\lambda\Delta P/L = p_c$
for the largest cluster which is not moving and assuming that the
separation of the channels, $l$, is the same as the largest non-moving
cluster diameter, i.e., $l \approx \lambda=p_c L/\Delta P$, one finds
$Q=ka^2W\Delta P^2/(\mu L^2p_c)$ or $Q\sim \Delta P^2$. Though this
scaling theory shows the quadratic dependence of $Q$ on $\Delta P$, it
does not contain any threshold pressure. Moreover, if we extend this
theory in 3 dimensions, the number of channels $n_l$ should be equal
to $W^2/l^2$ and following the same arguments we will find
$Q\sim\Delta P^3$ for 3D. This is contradictory to the mean field
theory \cite{sinhaepl12}, according to which the quadratic
relationship does not depend of the dimensionality of the network.
Secondly, the 3D experiments reported in Ref \cite{rassi11} showed
that, without considering the threshold pressure, the log-log plot of
$Q$ and $\Delta P$ shows a variation in the slope in the range $0.45$
to $0.3$. Those experiments were performed for a short range of
capillary numbers and were difficult to draw any definitive
conclusions from. It is therefore very important to perform extensive
experiments and numerical simulations of two-phase flow in 3D porous
medium to find the exact non-linear dependence of $Q$ and $\Delta P$
and to check whether the scaling exponent varies with dimensionality
of the system.

To our knowledge there is very little experimental work exploring
these flows using a three-dimensional pore network. In this article,
we present an extensive experimental and numerical study of the
steady-state two-phase flow of immiscible Newtonian fluids. In the
following, first we will present the experimental study of
steady-state two-phase flow of air and deionized water. This work
utilizes a three-dimensional porous medium, resulting in a more
chaotic flow and increased fluctuations around a mean
value. Significant work was done to collect high fidelity data and to
accurately obtain the average differential pressures in the
steady-state. We will then present our numerical work, describing the
network model of two-phase flow in three dimensional reconstructed
pore networks. From our experimental and computational results, we
will show that there exists a global threshold pressure $P_t$, below
which there is no flow through the system. The results are highly
significant in understanding the non-Newtonian two-phase flow behavior
of Newtonian fluids due to the dominant capillary forces in a
disordered system.


\section{Experimental Setup}
\label{sec:exptsetup}

\begin{figure}
  \centerline{\hfill\includegraphics[width=0.5\textwidth,clip]{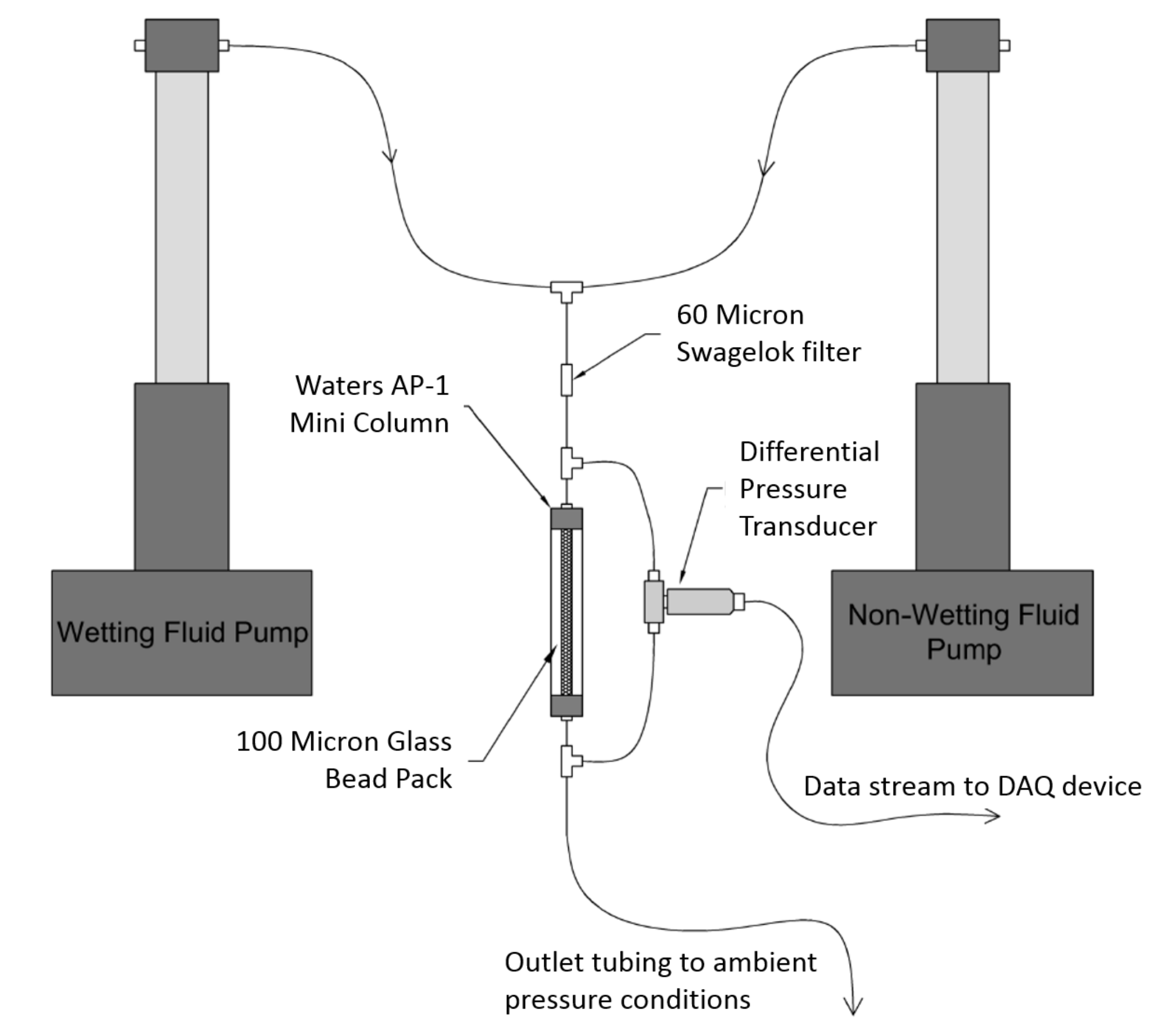}\hfill}
  \caption{\label{expt-setup} The experimental setup of two-phase flow
    of deionized water as the wetting fluid and air as the non-wetting
    fluid through the column of borosilicate glass beads.}
\end{figure}

Experiments were performed on a three-dimensional porous medium
confined in a vertical column. The porous medium was comprised of
randomly distributed borosilicate glass beads with diameter $d=98\pm 8
\upmu\text{m}$ and porosity $\phi=0.44$ (Cospheric LLC, Santa Barbara,
CA). These micro-spheres were wet-filled and well packed into a Waters
AP-1 Mini Chromatography Column with inner diameter $D=5\text{mm}$ and
length $L=70\text{mm}$. Filters at the inlet and outlet of the column
consisted of a bed of glass microfibers and fine plastic mesh. The
mesh allowed for uninhibited fluid flow, yet prevented the loss of
glass beads. The filters aided in mixing the fluids to further
homogenize the two phases before entering the porous media. This was
done to emulate the alternating inlet boundary conditions as used in
other experiments \cite{tallakstad09b} or simulations
\cite{sinhaepl12} and to minimize the slug-like flow of alternating
fluids. The two immiscible fluids used in this study were deionized
water as the wetting fluid, and air as the non-wetting fluid, having
respective dynamic viscosities $\mu_\text{w} = 1.002\text{Pa.s}$ and
$\mu_\text{nw} = 1.84\times 10^{-5}\text{Pa.s}$. This provides the
viscosity ratio $M=\mu_\text{nw}/\mu_\text{w}$ equal to $1.8363\times
10^{-5}$. The air-water surface tension at room temperature is around
$0.073$ N/m. The bond number comparing gravitational effects to the
surface tension, $E_o=\Delta\rho g d^2/\gamma$ where $\Delta\rho$ is
the difference in density of the fluids and $g$ the gravitational
acceleration, is around 0.001. If we assume the size of an air bubble
in the system is of the order of the column diameter $D=5$ mm, the
ratio between the buoyancy and the capillary force is $(D/d)E_0\approx
0.05$ making it possible to ignore gravitation in the analysis.

Two Teledyne Isco Model 500D syringe pumps were used to control the
bulk flow rates of each respective fluid, which were then combined in
a junction just above the column inlet. Polyetheretherketone (PEEK)
tubing (Sigma-Aldrich Co.) and stainless steel fittings (Swagelok)
comprised the flow system, and a sintered stainless steel
$60\upmu\text{m}$ filter element was used before the inlet to collect
any small particulates in the inlet fluids and homogenize the
phases. An Omega PX409 differential pressure transducer measured the
pressure gradient over the column of glass micro-spheres. Pressure
data was acquired using National Instrument’s LabVIEW at a sampling
rate of $1000$ samples per second, although sample compression was
used to average $1000$ samples into one data point, yielding one
binned data point per second. A schematic of the experimental setup
can be seen in Figure \ref{expt-setup}.


\section{Experimental Results and Discussions}
\label{sec:exptres}
Our experimental study is focused on measuring the scaling coefficient
$\alpha$ related to the scaling of the flow rate and the excess
pressure drop, $(\Delta P-P_t)\sim\text{Ca}^\alpha$, which has been
shown to be independent of fractional flow, saturation and viscosity
ratio \cite{tallakstad09b,sinhaepl12}. Our experiments use non-wetting
fractional flow $F_\text{nw}=0.5$ allowing for the volume of both
syringe pumps to be fully utilized. All experimental parameters were
kept constant except the flow rates of the fluids, so the overall bulk
flow rate was varied initially in order to reach the desired capillary
number. This paper focuses on experiments carried out at capillary
numbers $10^{-4}$ to $10^{-5.4}$. Before each experiment, the bead
pack was initially saturated with water (no air flow) which provided
consistency between experiments. The experiments were conducted by
simultaneously starting the flow of both water and air at the same
flow rate, resulting in the desired non-wetting fractional flow of
$F_\text{w}=0.5$. The total flow rate $Q$ for any given capillary
number $\text{Ca}$ is obtained from the fluid velocity $v$ given by
$v=\gamma\text{Ca}/\mu$, and then multiplying it with the effective
cross-section $A$, $Q=Av$. Here $\mu$ is the effective viscosity at
room temperature. The effective cross-sectional area $A$ is obtained
from the total cross-sectional area of the bead pack multiplied with
its porosity ($\phi=0.44$). The volumetric flow rate for each fluid
(air and water) is then half of the calculated total flow rate
$Q$. Due to the compressibility of the air, the phases likely competed
at the junction, causing some degree alternating injections of water
and air. Again, the filter was utilized as a means to reduce this
phenomenon and create a more homogenized flow into the bead pack. The
non-wetting saturation of the bead pack increased as air became
randomly distributed throughout the porous medium. The confining
column was borosilicate glass which provided for basic visual
observations of the bead pack saturation.

\begin{figure}
  \centerline{\hfill
    \includegraphics[width=0.32\textwidth,clip]{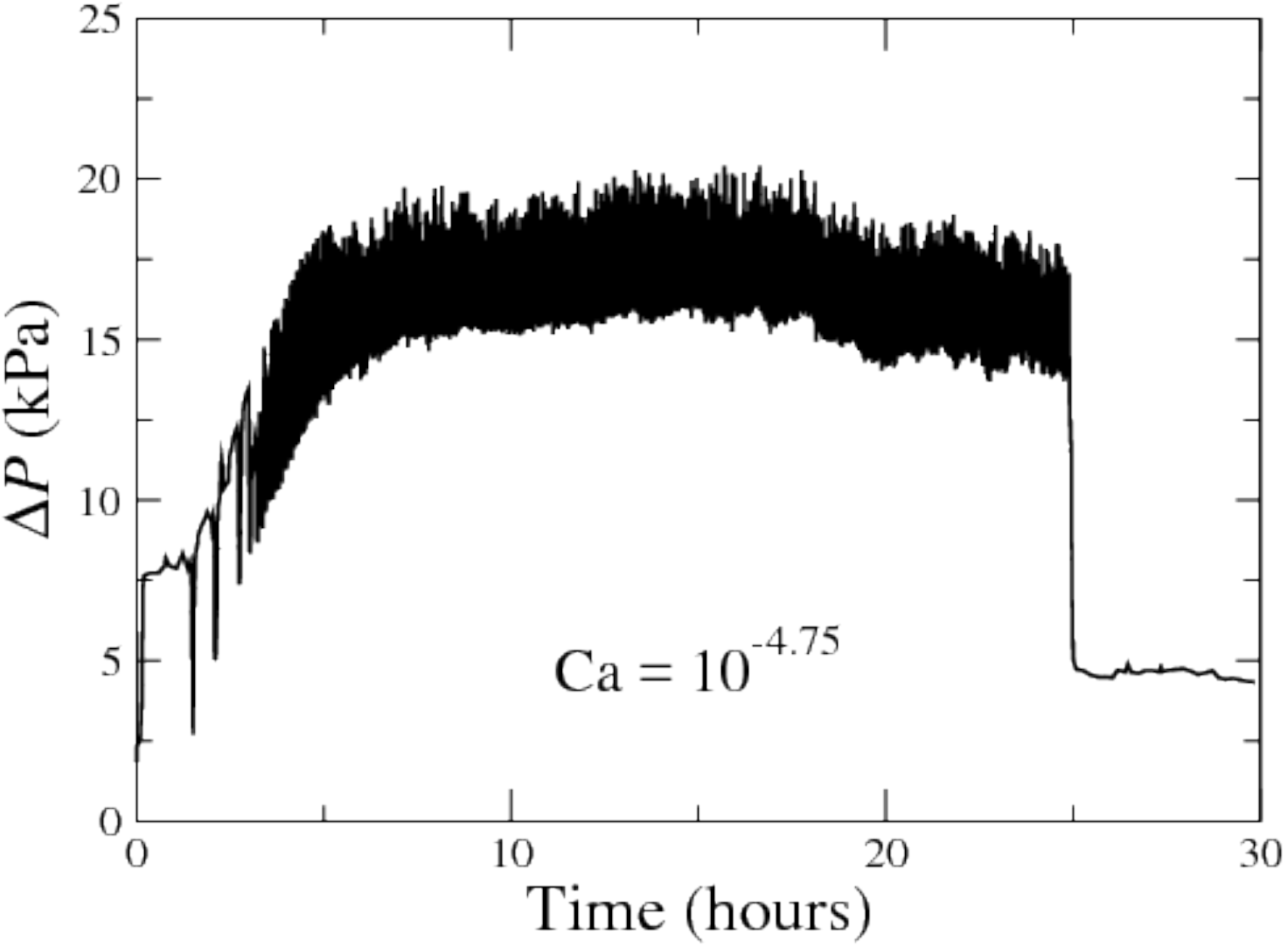}\hfill
    \includegraphics[width=0.32\textwidth,clip]{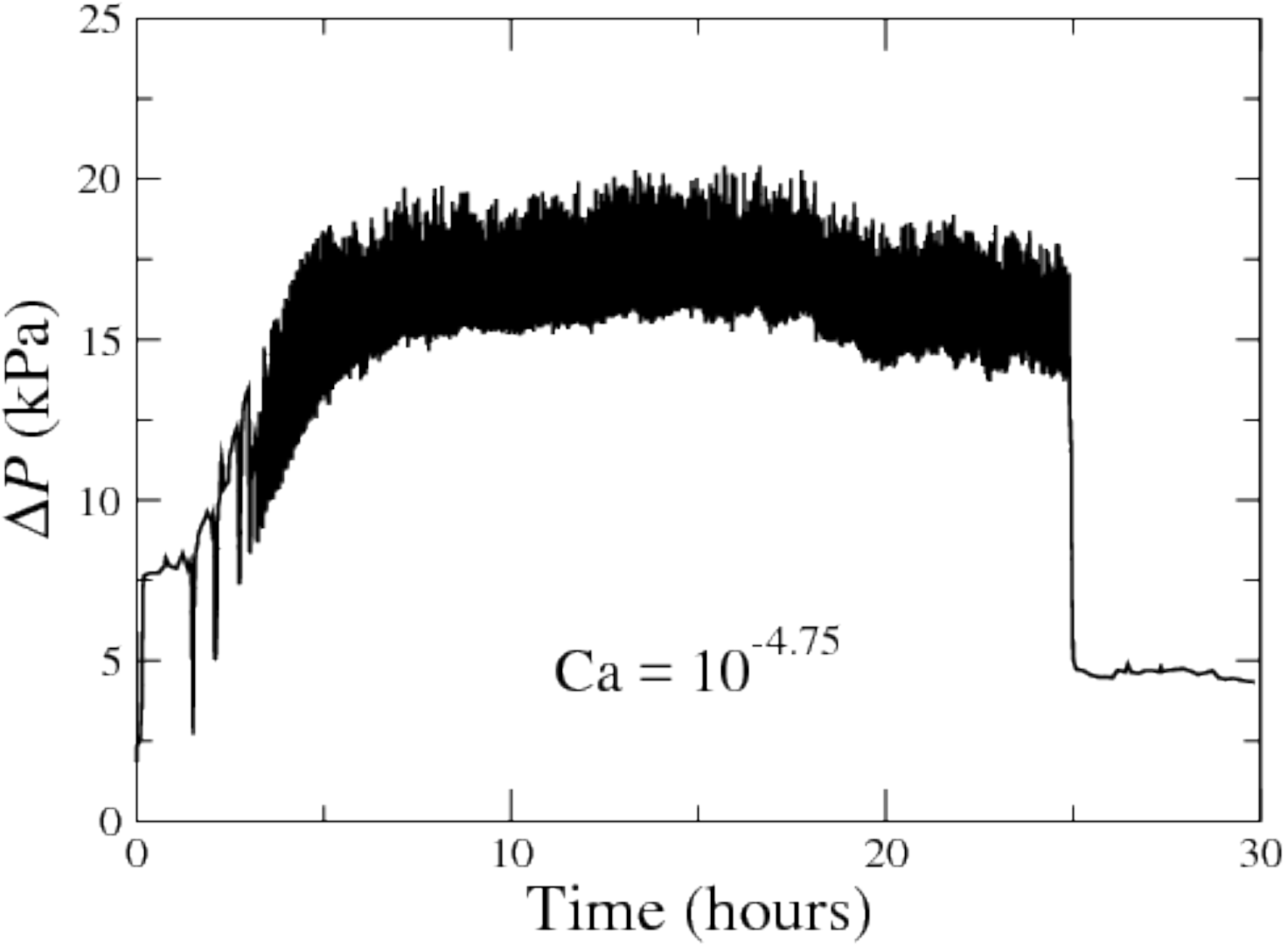}\hfill
    \includegraphics[width=0.32\textwidth,clip]{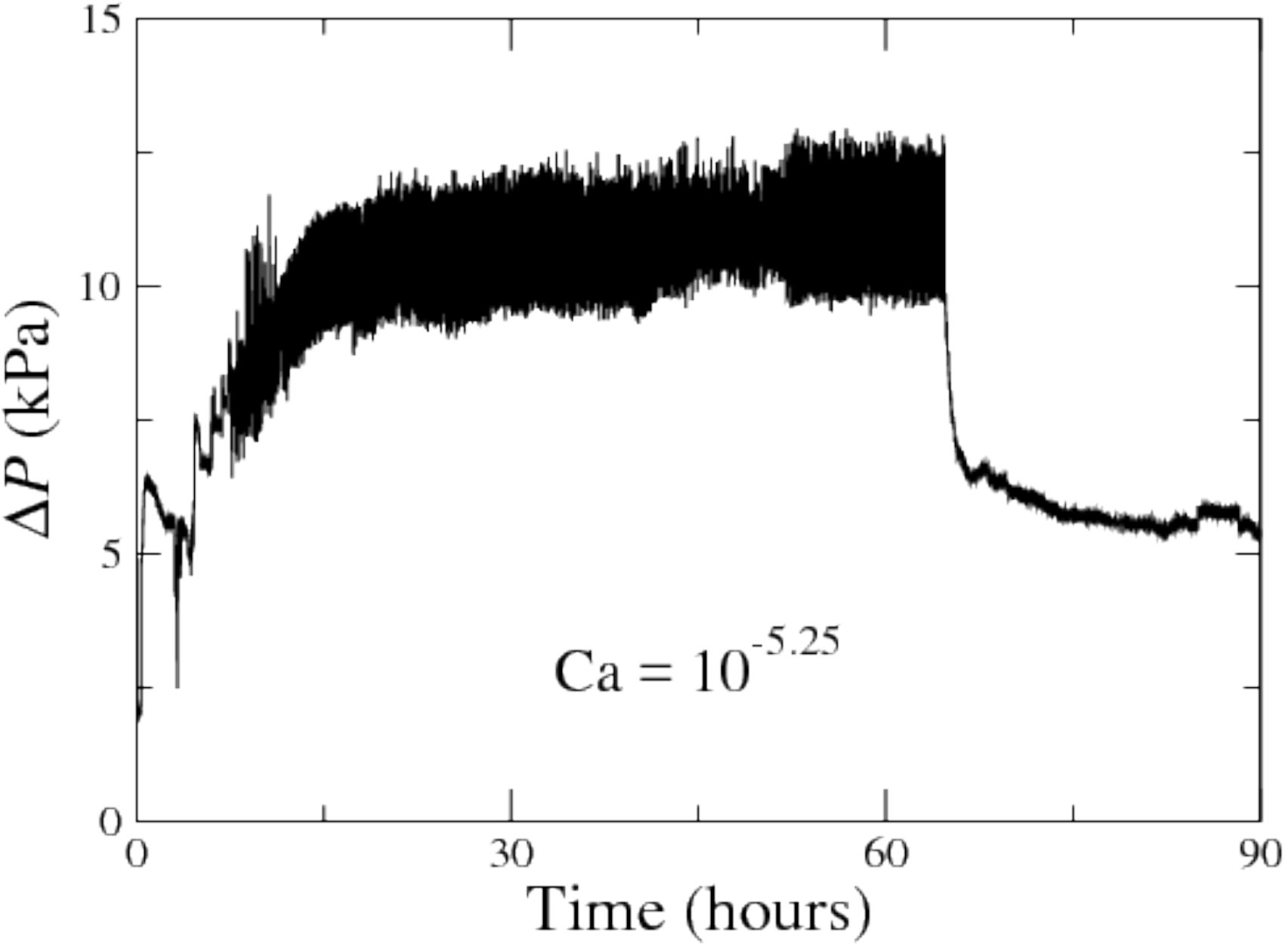}\hfill}
  \caption{\label{expt-pres}Three sample pressure plots of the time
    evolution of the experiment are shown to demonstrate the chaotic
    and transient nature of the two-phase flow. Plots from the high Ca
    regime (a), transition point (b), and low Ca regime (c) illustrate
    the characteristic pressure behavior observed.}
\end{figure}

In Figure \ref{expt-pres} pressure profiles are plotted over the
duration of experiments at respective Ca. The pressure gradient during
the transient regime followed a chaotic, increasing
trajectory. Reaching the steady state regime, the pressure fluctuated
around an average differential pressure $\Delta P$. Capillary forces
between the competing phases manifest in large fluctuations of $\Delta
P$ even at lower Ca values. The $\Delta P$ fluctuations maintain the
same overall magnitude and therefore more pronounced in the low
$\text{Ca}$ regime when the absolute value of $\Delta P$ is
lower. Upon stopping the pumps we have $Q=0$, however, the pressure
gradient over the column fell to a non-zero value which we identify as
the threshold pressure $P_t$. We were able to obtain experimental
values for $P_t$ that slightly differed between experiments and took
the average of these values as a global threshold pressure for the
column, $P_t=5.39$kPa.

As expected, the pressure gradient versus capillary number
relationship scaled nearly linearly with $\alpha=0.99$ for capillary
numbers $10^{-4.75}$ and higher. At low capillary numbers, it was
shown that the pressure gradient relationship scaled non-linearly with
$\alpha=0.46$. These values support the power scaling theory from
Sinha {\it et al.} and demonstrate that incorporating $P_t$ is crucial
to obtain a consistent scaling factor of $\alpha = 0.46$. There was a
distinct change from the Newtonian to non-Newtonian flow regime at
$\text{Ca}=10^{-4.75}$. This transition point was far from the value
found by similar experimental and numerical studies performed in
two-dimensions \cite{tallakstad09,sinhaepl12}.

\begin{figure}
  \centerline{\hfill\includegraphics[width=0.5\textwidth,clip]{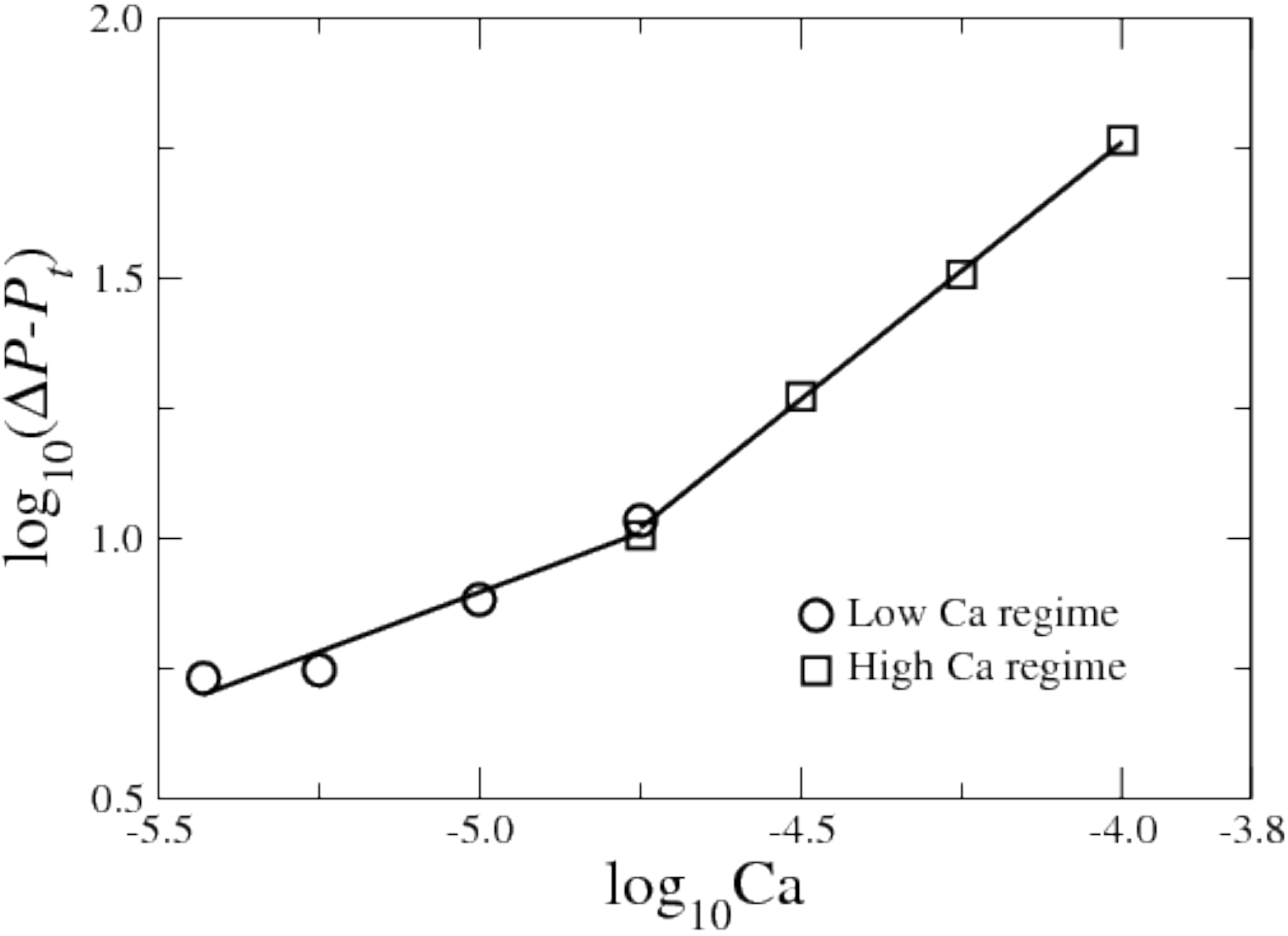}\hfill}
  \caption{\label{expt-scaling} Plot of excess pressure drop ($\Delta
    P-P_t$) as a function of the capillary number ($\text{Ca}$) for
    the nonlinear and linear flow regimes obtained from the
    experiment. Each data point represents the averaged steady state
    pressure gradient for the respective $\text{Ca}$. The two scaling
    exponents for the low and the high $\text{Ca}$ regimes obtained
    from the slopes are $0.46\pm 0.05$ and $0.99\pm 0.02$
    respectively.}
\end{figure}

During initial experimental runs, we observed an ageing effect in our
porous medium. The magnitude of $\Delta P$ required to maintain a
specific two-phase flow rate increased if a time span of several weeks
elapsed between measurements, yet all data collected within a short
time frame would show a scaling of $0.4\mbox{--}0.5$ in the low
$\text{Ca}$ regime, or $0.95\mbox{--}1.05$ in the high $\text{Ca}$
regime. Some of this was explained initially by biofouling of the
inline filter and mesh, which was avoided by adding a small amount of
biocide to the flowing fluid. However, in two-phase flow experiments
with a Hele-Shaw cell, Aursj{\o} {\it et al.} observed this ageing
effect phenomena and theorized that it was due to wetting effects
\cite{aursj14}. Similarly, we hypothesize wetting effects within the
column of glass micro-spheres may have contributed to the change of
the non-wetting saturation $S_\text{nw}$, resulting in a different
$\Delta P$ for a given two-phase flow rate. This theory is
corroborated by the simulations which show $\Delta P$ is a function of
$S_\text{nw}$ (see Figure \ref{ps-steady}). We would like to note that
the final data set was performed over the smallest time span possible
in order to avoid all ageing effects.


\section{Network Model for Two-Phase Flow in Reconstructed Pore Network}
\label{sec:model}

\begin{figure}
  \centerline{\hfill
    \includegraphics[width = 0.6\textwidth,clip]{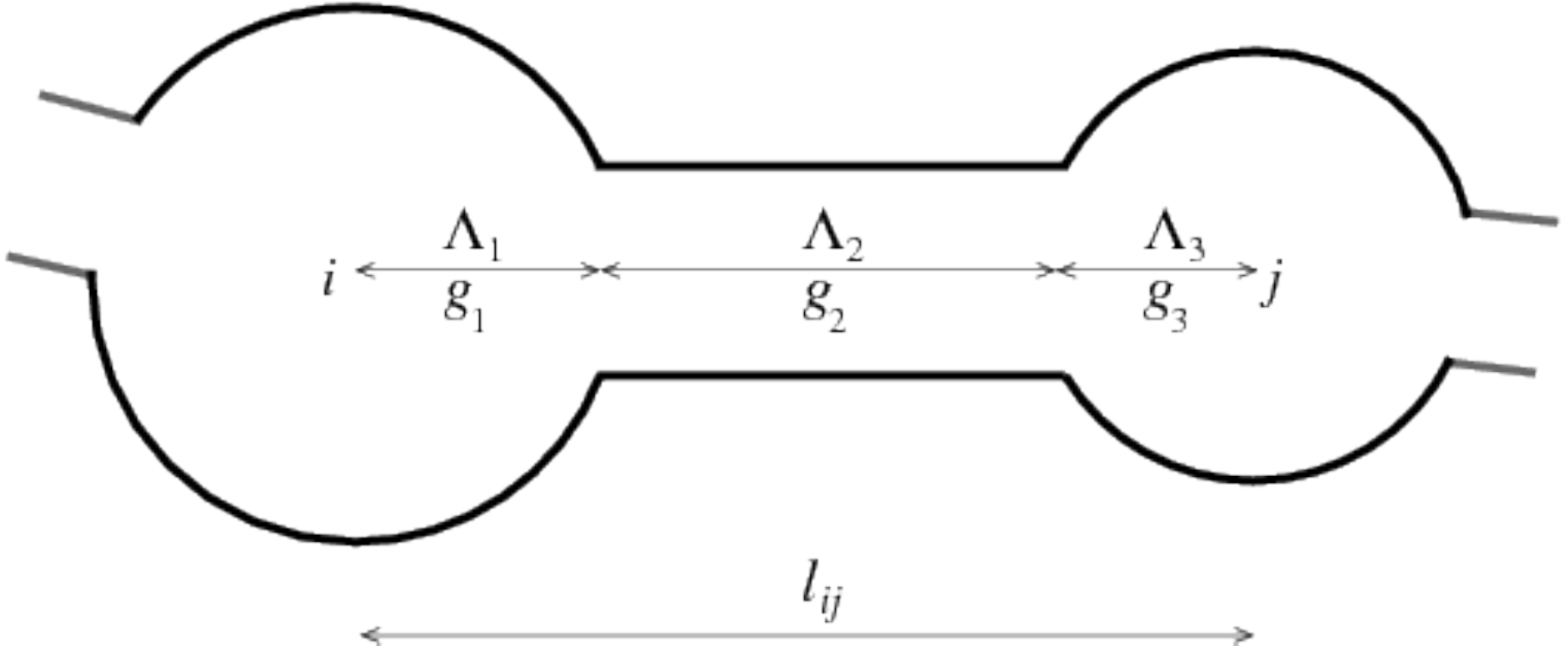}\hfill
    \includegraphics[width = 0.3\textwidth,clip]{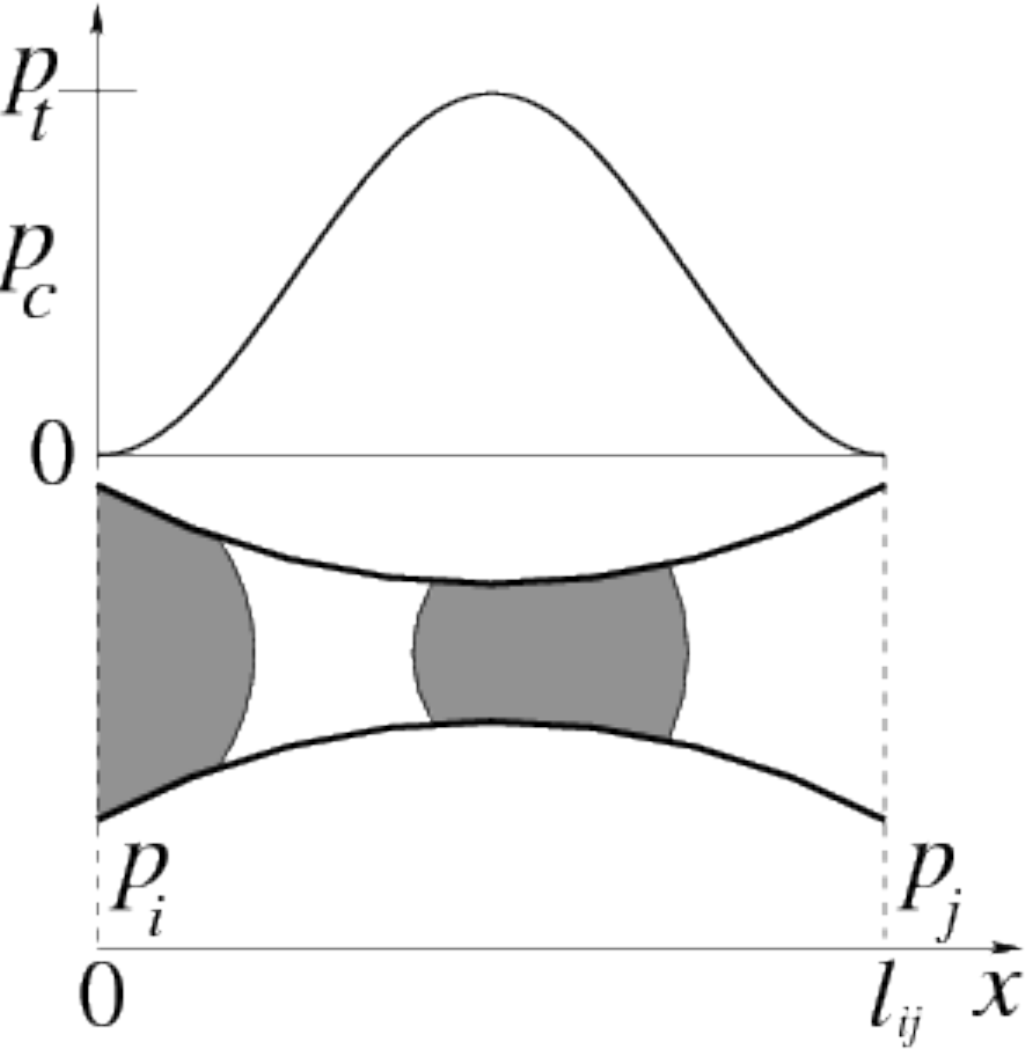}\hfill}
  \centerline{\hfill \hfill (a) \hfill \hfill \hfill (b) \hfill}
\caption{\label{model01} The schematic of one link between the two
  nodes $i$ and $j$ is shown in (a). The pore space of each link is
  divided into three pore-parts, two pore-bodies at two ends and one
  pore-throat in between. The total length ($l_{ij}$) of the link is
  equal to $\Lambda_1+\Lambda_2+\Lambda_3$, the sum of each
  pore-part. The presence of multiple interfaces between the wetting
  (white) and non-wetting (gray) fluids in a link is shown in the
  bottom of (b) and the variation of capillary pressure $p_c(x)$ as a
  function of the interface position ($x$) for each interface is shown
  in the top of (b). $p_c=0$ at the two ends of the link and is
  maximum, equal to the threshold pressure $p_t$, at the middle of the
  tube. $p_t=4\gamma\cos\theta/r_t$ is the minimum pressure required
  for the non-wetting fluid to invade the pore.}
\end{figure}

Network models of multiphase flow are useful to understand the
macroscopic properties of large pore networks relating the underlying
pore-scale physics of a porous material \cite{hass12}. A few decades
ago, medical micro-CT (micro computed X-ray tomography) scanners were
adopted to scan geological samples \cite{hurst84} and since then there
has been a revolution in the new scanning techniques to characterize
the micro-structures of porous media \cite{dunsmoir91,bbdgimpp13}. To
use these pore structures obtained from the 3D images of the pore
space as the input to the network models, simplified networks
consisting of pores and throats are reconstructed using different
approaches. One approach is to use a statistical model \cite{blunt05}
where different statistical properties like the porosity distribution,
correlation function and linear path function are estimated from the
images of 2D thin sections of the sample. Random 3D networks are then
generated with the same statistical properties obtained from the
images. This method is often questioned and observed to match poorly
with the original sample due to the loss of long range geometric
connectivity \cite{Hilfer00,jiao09,oren02}. Another method of
reconstruction is by {\it process-based} models, where 2D thin section
images are analyzed to measure the grain size distribution and other
petrophysical properties and then the packing of the grains are
simulated following different geological processes, such as
sedimentation, compaction, rearrangement and diagenesis
\cite{oren02,oren03}. These models show good results for samples where
sedimentary processes are involved, such as sandstones. However in the
case of systems having complex sedimentary and diagenetic history or
systems with heterogeneity -- such as carbonates -- network
reconstruction using process based models are difficult. In such
cases, to extract the pore networks from any generic 3D image of an
arbitrary porous medium, different voxel based models such as medial
axis based methods and maximum ball methods are used
\cite{hudongthesis07,blunt09}. More details on the different network
reconstruction methods can be obtained in the respective
references. In the present study we use three different networks
reconstructed from (A) Berea sandstone, (B) sandpack and (C) a
sandstone (``sandstone 9" in \cite{hudongthesis07}).  Networks A and B
were reconstructed using the process-based models \cite{oren02,oren03}
whereas C was reconstructed using a voxel based maximum ball algorithm
\cite{hudongthesis07,blunt09}. The physical dimension of the samples
A, B and C and the number of links and nodes of the corresponding
reconstructed networks are listed in Table \ref{netparam}.

Sample A and B have been used previously e.g. in Ramstad {\it et al.}
\cite{rho09} and T{\o}r{\aa} {\it et al.} \cite{toh12}
respectively. Samples C is described in \cite{hudongthesis07} and may
be found at \cite{bluntweb}.

The network extracted from a pore sample consists of links that are
connected at nodes. The number of links connected to a node is the
degree or the coordination number of that node. Each link is
associated with a set of parameters which characterize the pore space
of the original sample. A link has $3$ pore-parts, two pore bodies at
the end and one pore throat in between as shown in Figure
\ref{model01}. The cross-section of the pores are triangular in shape
and characterized by a shape factor $G$, defined as the ratio between
the effective cross-sectional area of the pore ($a$) and the square of
its circumference. The value of $G$ can vary in the range
$(0,\sqrt{3}/36]$ for triangular cross-section where the largest value
($\approx 0.048$) corresponds to an equilateral triangle. The
effective cross-sectional area can then be calculated from the
relation $a=r^2/(4G)$, where $r$ is the radius of the inscribed circle
in the pore-part \cite{mason91}. The network transports two immiscible
fluids, one is more wetting than the other with respect to the pore
wall. We consider that the wetting properties of the fluids are such
that there is no film flow in the system and the flow in the pores are
piston-like. The instantaneous local flow-rate $q_{ij}$ inside a link
between two nodes $i$ and $j$ follows the Washburn equation of
capillary flow \cite{washburn21,aker98},
\begin{equation}
  \label{washburneq}
  \displaystyle
  q_{ij}= \frac{g_{ij}}{l_{ij}}\left[p_j-p_i-\sum p_c(x)\right],
\end{equation}
where $p_i$ and $p_j$ are the local pressure drops at $i$th and $j$th
nodes. The mobility $g_{ij}$ of the link between the two nodes is
calculated from the harmonic average of the individual conductances
from each of three pore-parts of that link, given by,
\begin{equation}
  \label{mobavg}
  \displaystyle
  \frac{l_{ij}}{g_{ij}}=\frac{\Lambda_1}{g_1}+\frac{\Lambda_2}{g_2}+\frac{\Lambda_3}{g_3},
\end{equation}
where $\Lambda_{1,2,3}$ and $g_{1,2,3}$ are the lengths and
conductances of each pore-part respectively as shown in Figure
\ref{model01}. For a triangular cross-section, each individual term of
$g_{1,2,3}$ for each pore-part of a link is given by
\cite{langglois64,jia08},
\begin{equation}
  \label{mobpart}
  \displaystyle
  g = \frac{3r^2a}{20\mu_p},
\end{equation}
where $a$ and $r$ are respectively the effective area and the radius
of the inscribed circle in the pore part. $\mu_p$ is the
time-dependent saturation-weighted viscosity for the link given by
$\mu_p=\mu_\text{nw}s+\mu_\text{w}(1-s)$, where $s$ is the
instantaneous non-wetting saturation inside the link.

\begin{table}
  \centering
  \begin{tabular}{c|c|c|c}
    \hline
    Network & Physical dimension ($\text{mm}^3$) & Number of nodes & Number of links \\
    \hline
    A       & $1.8\times 1.8\times 1.8$          & $1163$          & $2274$ \\
    \hline
    B       & $4.5\times 1.5\times 1.5$          & $767$           & $1750$ \\
    \hline
    C       & $1.0194\times 1.0194\times 1.0194$ & $604$           & $1054$ \\
    \hline
    \end{tabular}
  \caption{\label{netparam} Physical dimensions of the porous media
    samples A, B and C and the number of links and nodes of the
    corresponding reconstructed networks.}
\end{table}

The capillary pressure $p_c(x)$ in equation \ref{washburneq} appears
due to the surface tension at the interfaces where $x\in[0,l_{ij}]$ is
the position of the interface. It acts as a barrier for the
non-wetting fluid to penetrate through the pores filled with the
wetting fluid and will be maximum at the narrowest part of the pore,
i.e. the pore throat. The pores are in between grains and the links
are therefore approximated as hourglass-shaped in the longitudinal
direction in terms of the capillary pressure. The functional
dependence of the capillary pressure on the interface position inside
such a pore is modeled by a modified form of Young Laplace equation
\cite{dullien92,aker98},
\begin{equation}
  \label{ylaplace}
  \displaystyle
  |p_c(x)| = \frac{2\gamma\cos\theta}{r_t}\left[1-\cos\left(\frac{2\pi x}{l}\right)\right],
\end{equation}
where $r_t$ denotes the throat radius which is the narrowest part of
the pore. $\gamma$ is the surface tension and $\theta$ is the contact
angle between the interface and the pore wall. In our simulations we
set $\gamma\cos\theta = 0.03$ N/m. The chosen form of $p_c(x)$
therefore provides the necessary $x$-dependence so that
$p_c(0)=p_c(l)=0$ and $\displaystyle\max_{x\in[0,l_{ij}]}|p_c(x)| =
|p_c(l_{ij}/2)|$. The summation over $p_c$ in equation
\ref{washburneq} runs over all the interfaces inside one link.

A constant volumetric flow rate $Q$ is generated through the
application of a pressure drop $\Delta P$ across the system.  $Q$ is
of course the same through any cross-section of the network.  Local
pressures ($p_i$) at each node are then determined by solving the set
of linear equations balancing the flow at each node using the
Kirchhoff equations, ensuring that the net flux in any node is
zero. This is done by solving the corresponding matrix inversion
problem using the conjugate gradient algorithm \cite{batr88}. When the
local node-pressures are known, the local flow rates $q_{ij}$ through
each link is calculated using equation \ref{washburneq}. This
determines the velocity of each interface inside any link. We choose
an adaptive time step $\Delta t$ in such a way that the displacement
of any meniscus does not exceed one-tenth of the length of the
corresponding link within that time. All the interfaces are moved
accordingly which changes the pressure distribution in the
network. The pressure at the nodes are then determined again using
conjugate gradient algorithm and the whole process is repeated. When
an interface reach at the end of a link, wetting and non-wetting
bubbles are snapped-off and new interfaces are created in the
neighboring links. The rules related to the interface dynamics used in
this 3D network model are similar to the 2D network model found in
\cite{aker98,knudsen02}. The in- and out-fluxes of the fluids from one
node to the neighboring links at any time step is determined from the
relative flow-rates of the links connected to that node. As this can
increase the number of interfaces in any link infinitely, we put a
limit in the maximum number of interfaces inside any link. When this
limit is exceeded, we merge the two nearest interfaces keeping the
volume of each fluid conserved. In the simulations reported in this
article, we have considered a maximum of $4$ interfaces in any link.

In order to reach steady state, we need to impose the periodic
boundary condition so that the fluid configurations that leave the
network from one side can enter from the opposite side and the flow
can continue for infinite time. However, the network here is irregular
and therefore the two opposite edge surfaces do not match each other
which is necessary to apply the periodic boundary condition. We solve
this problem by making a mirror copy of the network in the direction
of the overall flow and connected the copy with the original
network. The edges then match each other and periodic boundary
conditions are implemented in the direction of overall pressure
gradient. Notice that, this makes the system closed and the fluid
saturations do not change with time. The saturation is therefore a
control parameter here and we measure the fractional flow, whereas in
our experiments we control the fractional flow.

\section{Simulation results}
\label{sec:simres}

\begin{figure}
  \centerline{\hfill
    \includegraphics[width=0.10\textwidth,clip]{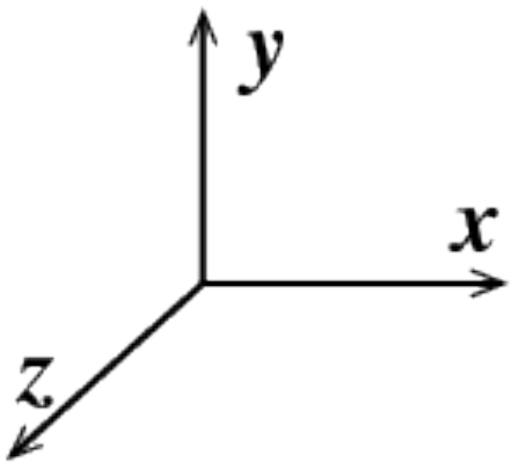}\hfill
    \includegraphics[width=0.28\textwidth,natwidth=1164,natheight=588,clip]{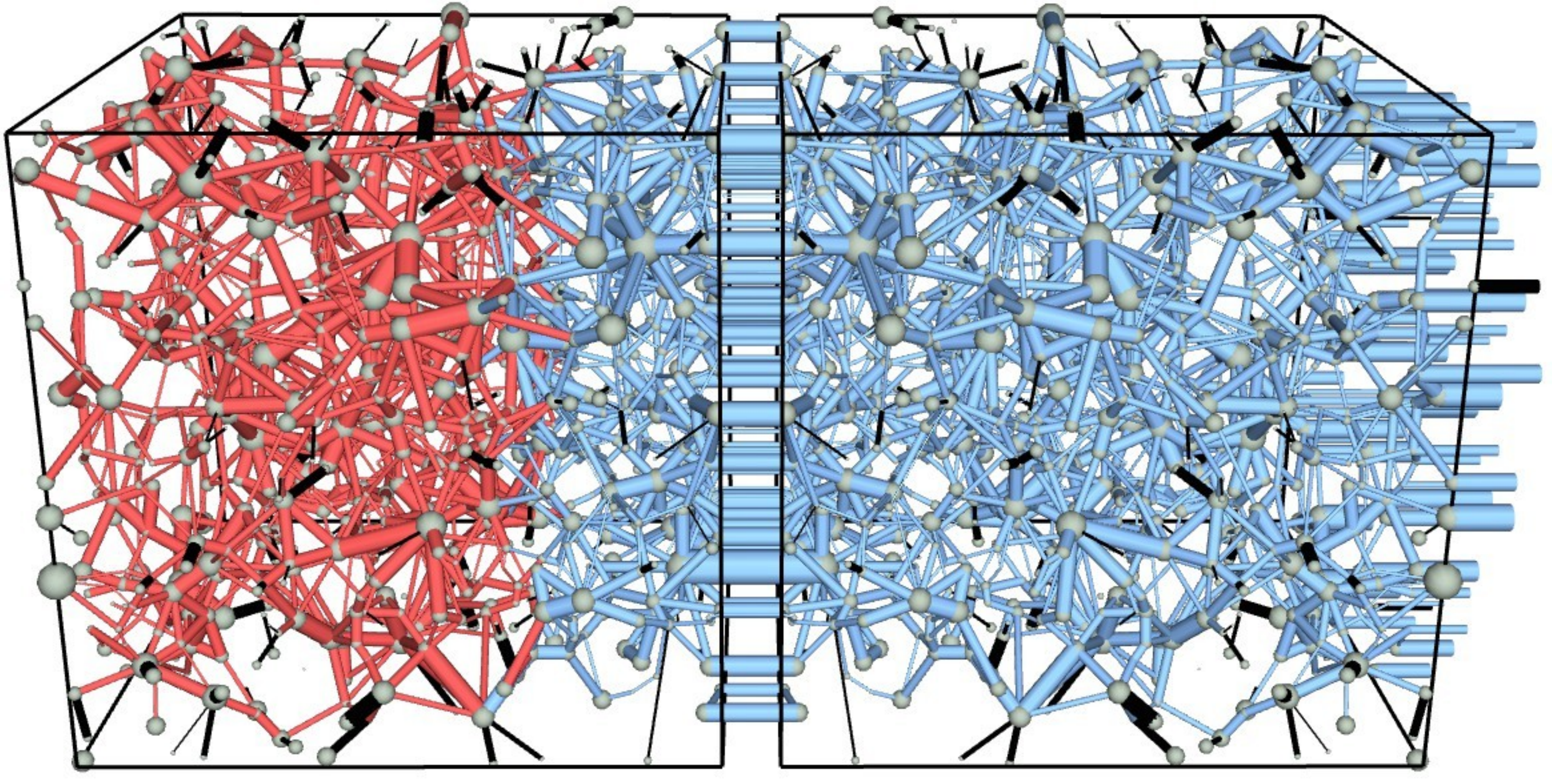}\hfill
    \includegraphics[width=0.28\textwidth,natwidth=1164,natheight=588,clip]{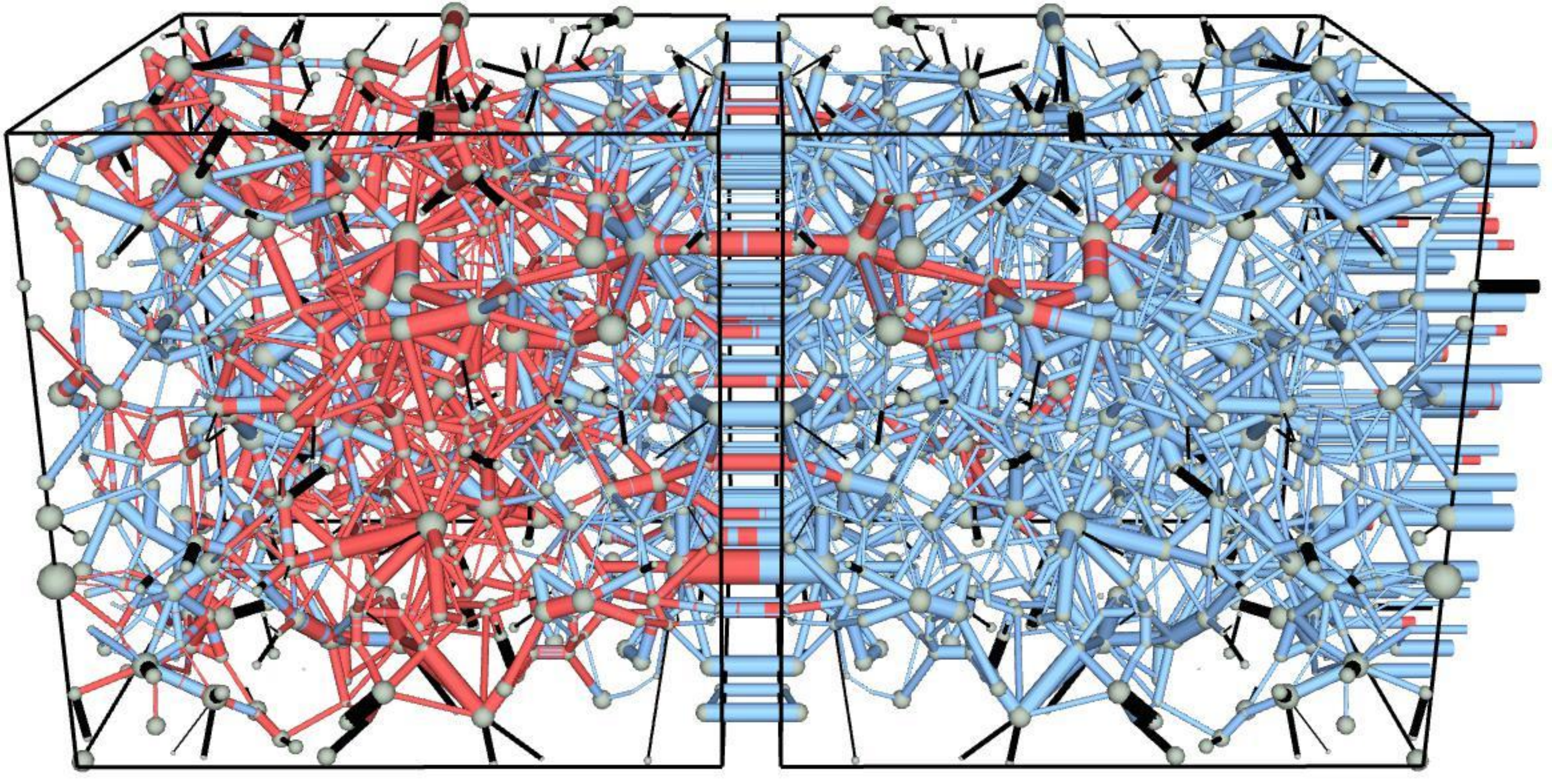}\hfill
    \includegraphics[width=0.28\textwidth,natwidth=1164,natheight=588,clip]{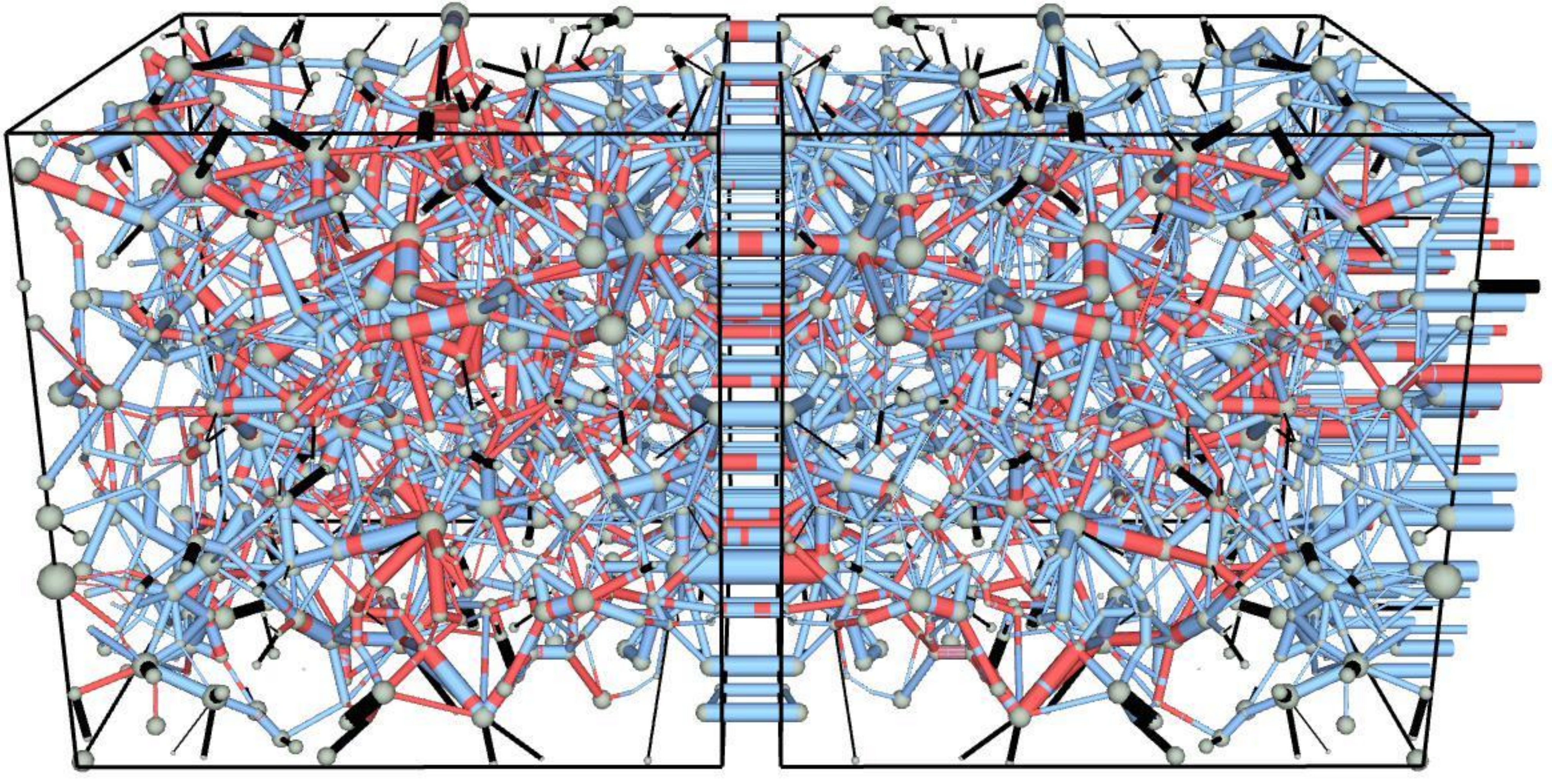}\hfill}
  \medskip
  \centerline{\hfill
    \includegraphics[width=0.32\textwidth,natwidth=1396,natheight=248,clip]{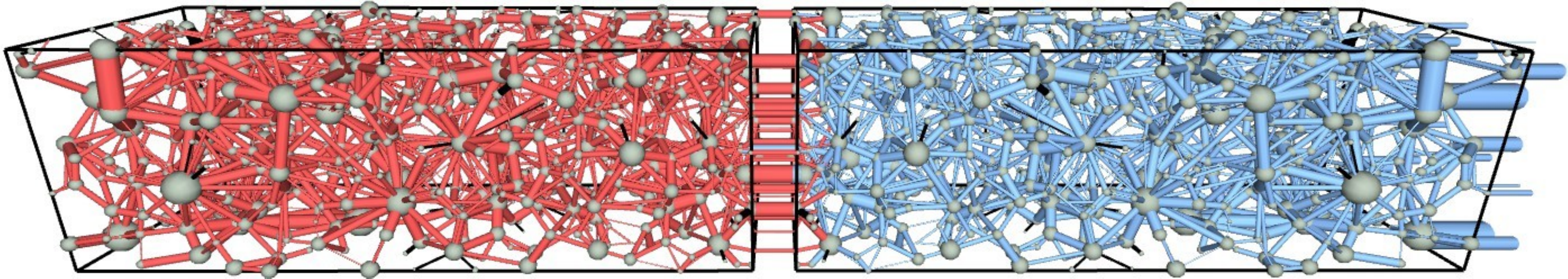}\hfill
    \includegraphics[width=0.32\textwidth,natwidth=1396,natheight=248,clip]{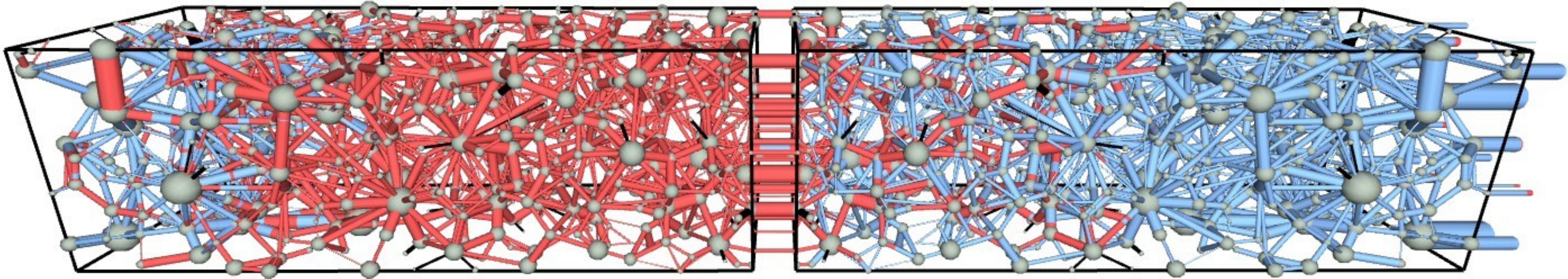}\hfill
    \includegraphics[width=0.32\textwidth,natwidth=1396,natheight=248,clip]{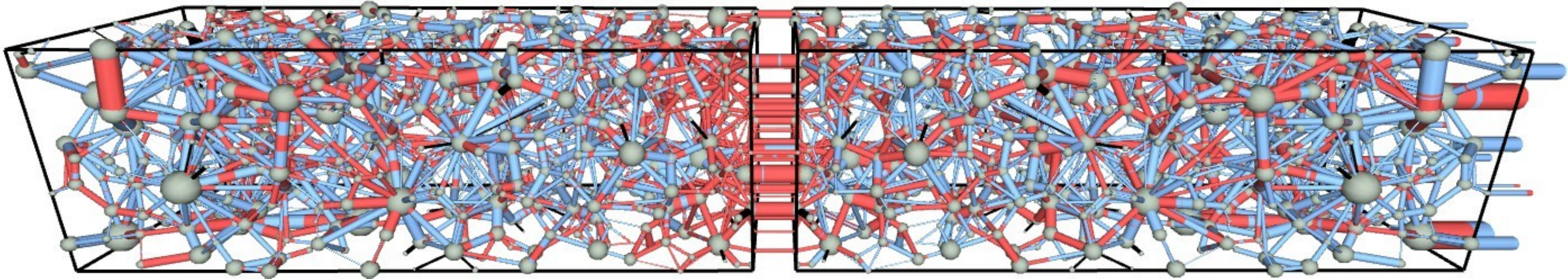}\hfill}
  \medskip
  \centerline{\hfill
    \includegraphics[width=0.32\textwidth,natwidth=1320,natheight=611,clip]{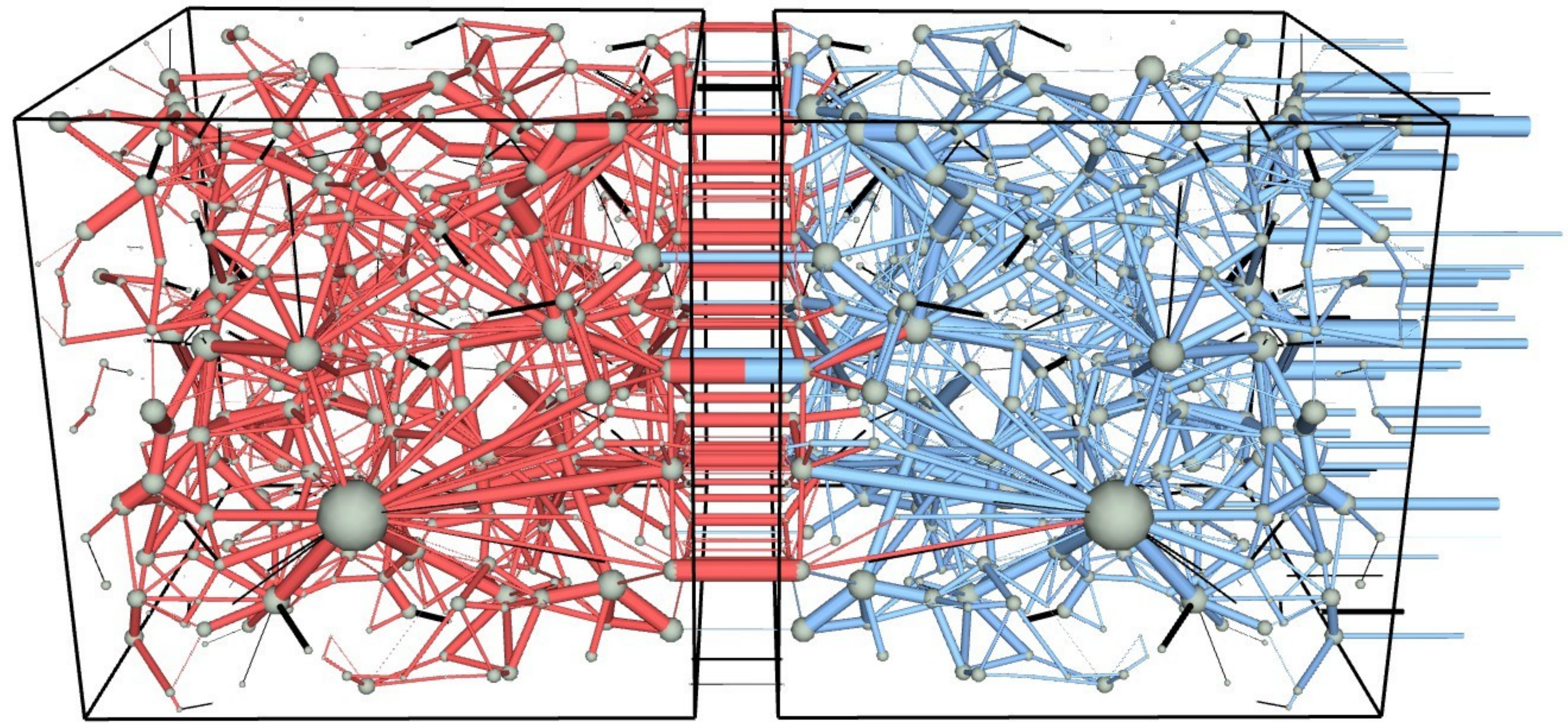}\hfill
    \includegraphics[width=0.32\textwidth,natwidth=1320,natheight=611,clip]{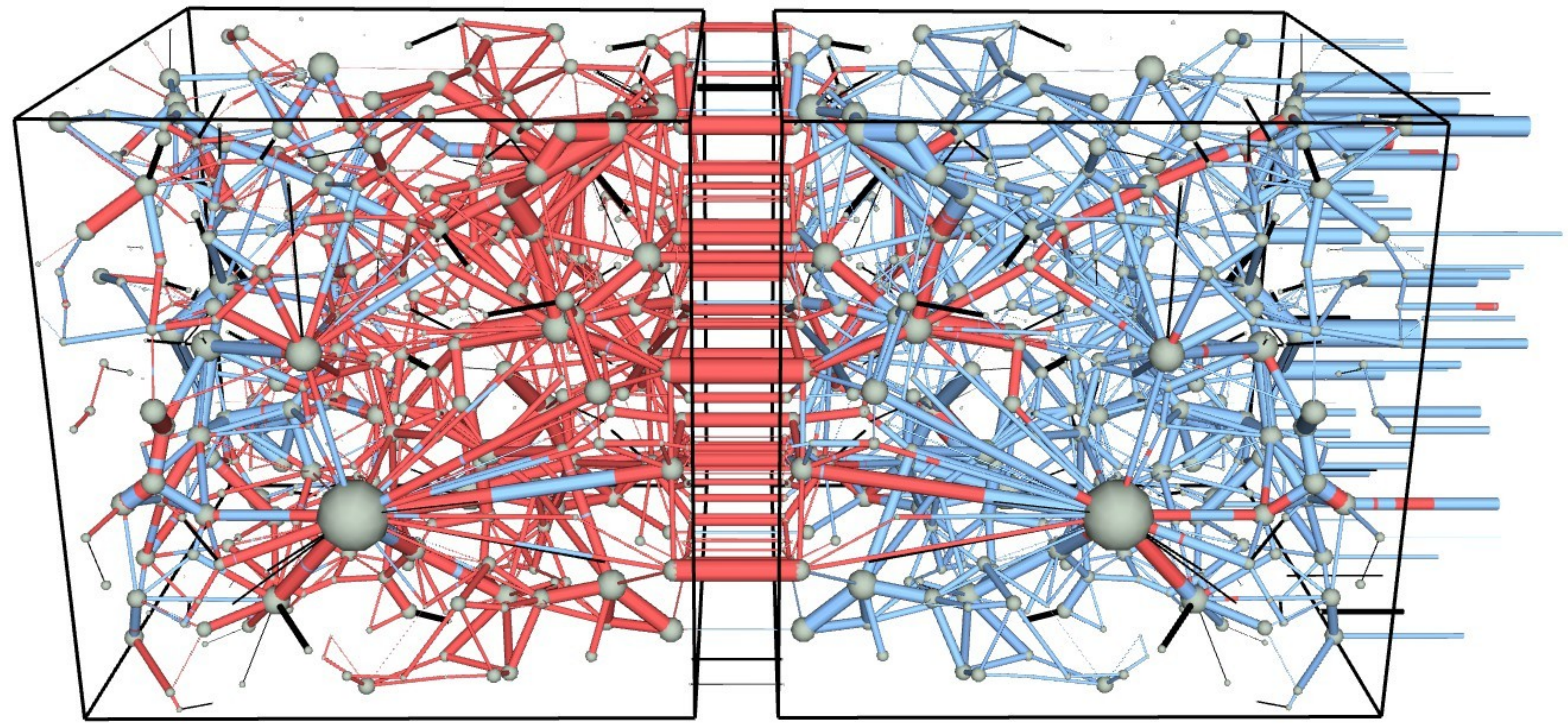}\hfill
    \includegraphics[width=0.32\textwidth,natwidth=1320,natheight=611,clip]{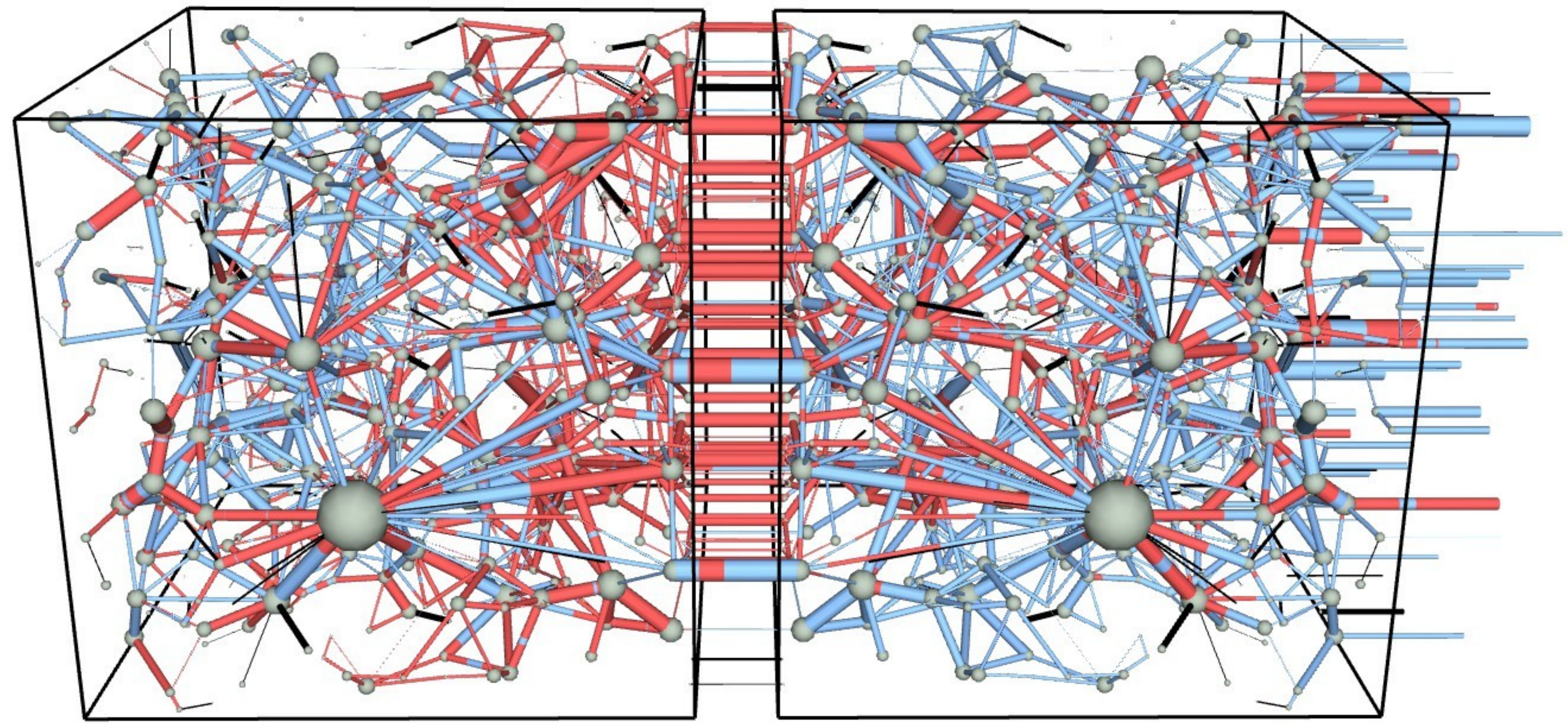}\hfill}
\caption{\label{snapshots} The time evolution of two-phase flow
  through the reconstructed networks where the wetting and the
  non-wetting fluids are colored by blue and red respectively. The few
  links in black are the dead ends which are connected only at one
  node and therefore removed from the network. The three images from
  left to right in each row respectively correspond to the initial
  condition of the system and after $0.1$ and $0.3$ pore volumes of
  fluids have passed. The three rows from top to bottom correspond to
  the samples A (berea), B (sandpack) and C (sandstone) respectively.
  The overall flow is in the positive $x$ direction. The periodic
  boundary condition is implemented in the same direction, by making a
  mirror image of the original reconstructed network and then
  connected together with the original. These two parts are shown by
  the two cuboids in the figures. Here, the system is initialized by
  filling the links with non-wetting fluid from $x=0$ until the
  required saturation is obtained and then filling the rest with the
  wetting fluid. In these figures, the non-wetting saturations are
  $0.3$ for sample A and $0.5$ for B and C.}
\end{figure}

Simulations are performed with constant flow rate $Q$, which sets the
capillary number $\text{Ca}$, the ratio of the viscous to the
capillary forces at the pore level, given by $\text{Ca}=Q\mu/(\gamma
A)$. Here $A$ is the cross-sectional area of the pore space, $\mu$ is
the saturation weighted viscosity and $\gamma$ is the surface tension
between the two fluids. Here we have considered two viscosity ratios,
$M=1$ where $\mu_\text{nw} = \mu_\text{w} = 0.1\text{Pa.s}$ and
$M=0.1$ where $\mu_\text{nw} = 10^{-3}\text{Pa.s}$ and
$\mu_\text{w}=10^{-2}\text{Pa.s}$. For a reservoir, the oil-water
viscosity ratio can vary in a wide range depending on the type of the
oil and the temperature \cite{barillas08}. Initial transients during
the simulation at $\text{Ca}=0.01$ and $M=1$ are shown in Figure
\ref{snapshots} where the wetting and non-wetting fluids are colored
by blue and red respectively. There is a few links marked by
black. These are those links which are connected only with one node
and will act as a dead end. The three rows from top to bottom
correspond to the three networks A, B and C respectively and the three
snapshots from left to right for each network are taken at three
different time steps - at the beginning of the simulation and after
$0.1$ and $0.3$ pore-volumes of fluid have passed. The overall flow is
in the positive $x$ direction. There are two cuboids in each network,
the one inside the left cuboid is the original reconstructed network
and in the right one is the mirror copy of that, which is done in
order to implement the periodic boundary condition as discussed
earlier. We prepared the initial system by filling it sequentially by
two fluids with necessary saturation so that the network is segregated
into one part of non-wetting (red) fluid and one part of wetting
(blue) fluid as shown in Figure \ref{snapshots}. Here
$S_\text{nw}=0.3$ for A and $0.5$ for B and C. When the simulation
starts, the non-wetting fluid starts invading the wetting fluid and
depending on the capillary number and viscosity ratio it will either
start viscous fingering for high $\text{Ca}$ or the capillary
fingering when the capillary forces dominate. On the other hand, the
wetting fluid also enters the system from left due to the periodic
boundary and pushes the non-wetting fluid. This displacement of
wetting fluid into the non-wetting is favorable and the displacement
should be piston-like or stable displacement. Though the network is
small here, one can still observe some signatures of these two
different types of fluid displacements in the initial transients -- in
the right part there are capillary fingers of the red fluid into the
blue whereas in the left the blue fluid displaces the red more
uniformly with compact propagating front. (More detailed time
evolutions of the fluid displacements are presented in the animations
in the supplementary material.)

\begin{figure}
  \centerline{\hfill
    \includegraphics[width=0.33\textwidth,clip]{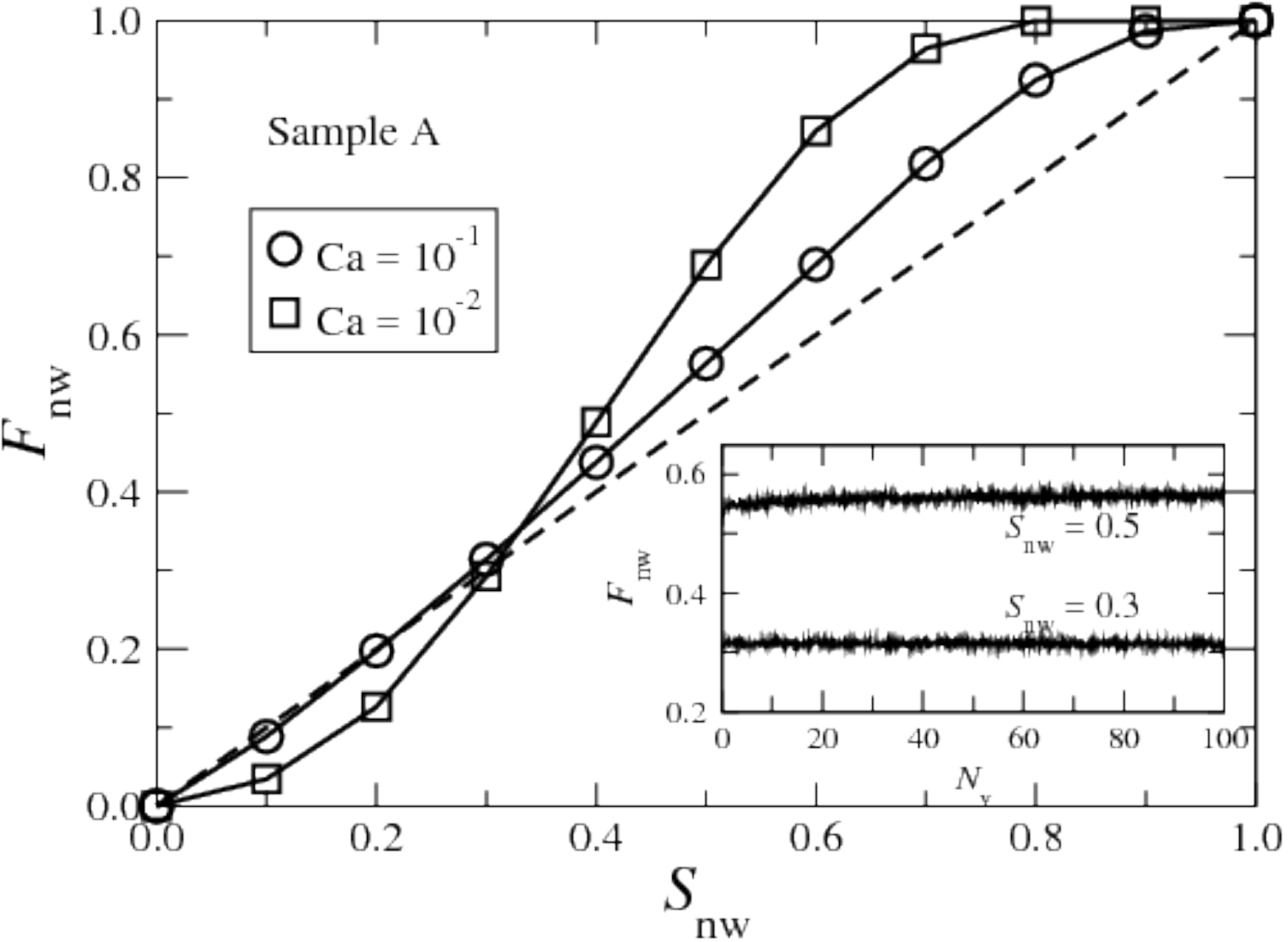}\hfill
    \includegraphics[width=0.33\textwidth,clip]{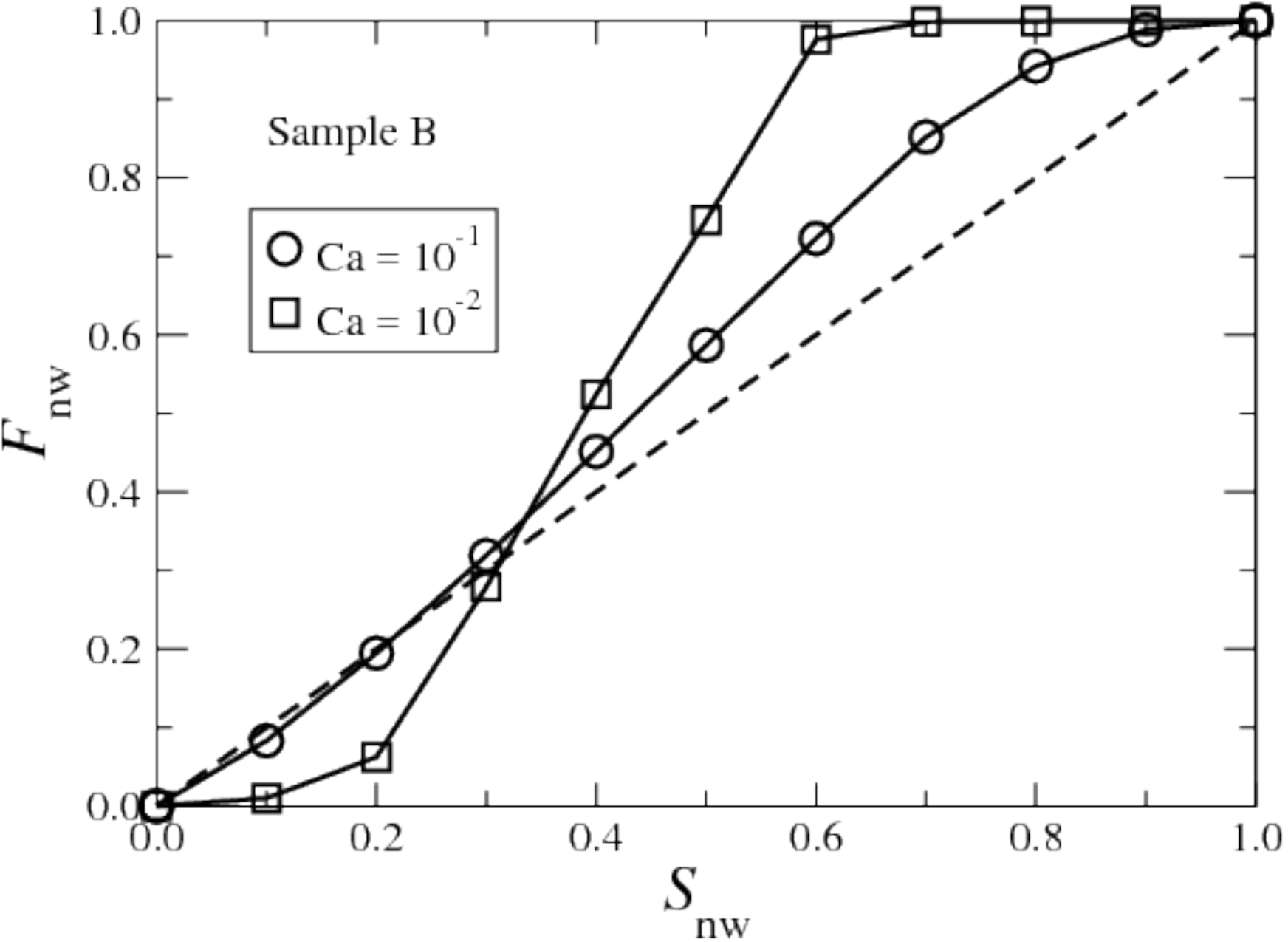}\hfill
    \includegraphics[width=0.33\textwidth,clip]{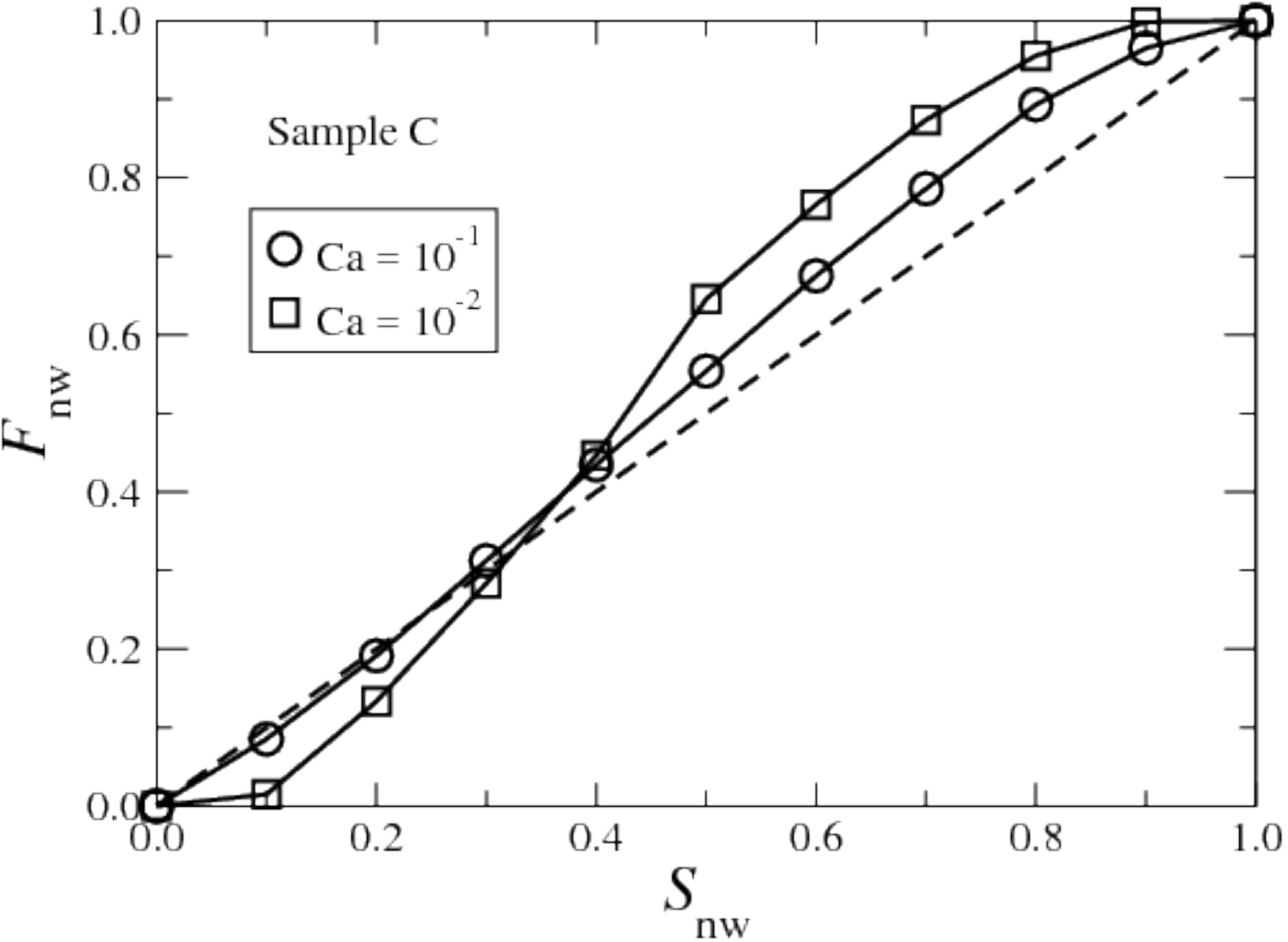}\hfill}
  \caption{\label{fs-steady}Plots of non-wetting fractional flow
    ($F_\text{nw}$) in the steady state as a function of non-wetting
    saturation ($S_\text{nw}$) for the three different networks. The
    dashed diagonal straight lines correspond to
    $F_\text{nw}=S_\text{nw}$, a system of miscible fluids will follow
    that line. The results for two different capillary numbers,
    $\text{Ca} = 10^{-1}$ and $10^{-2}$, are shown. Notice that the
    curves approach the diagonal straight line for the higher value of
    $\text{Ca}$. In the inset for sample A, we plot the fractional
    flow as a function of pore-volumes ($N_\text{v}$) of fluids
    passed, where $F_\text{nw}$ fluctuates around an average value in
    the steady-state.}
\end{figure}

\begin{figure}
  \centerline{\hfill
    \includegraphics[width=0.32\textwidth,clip]{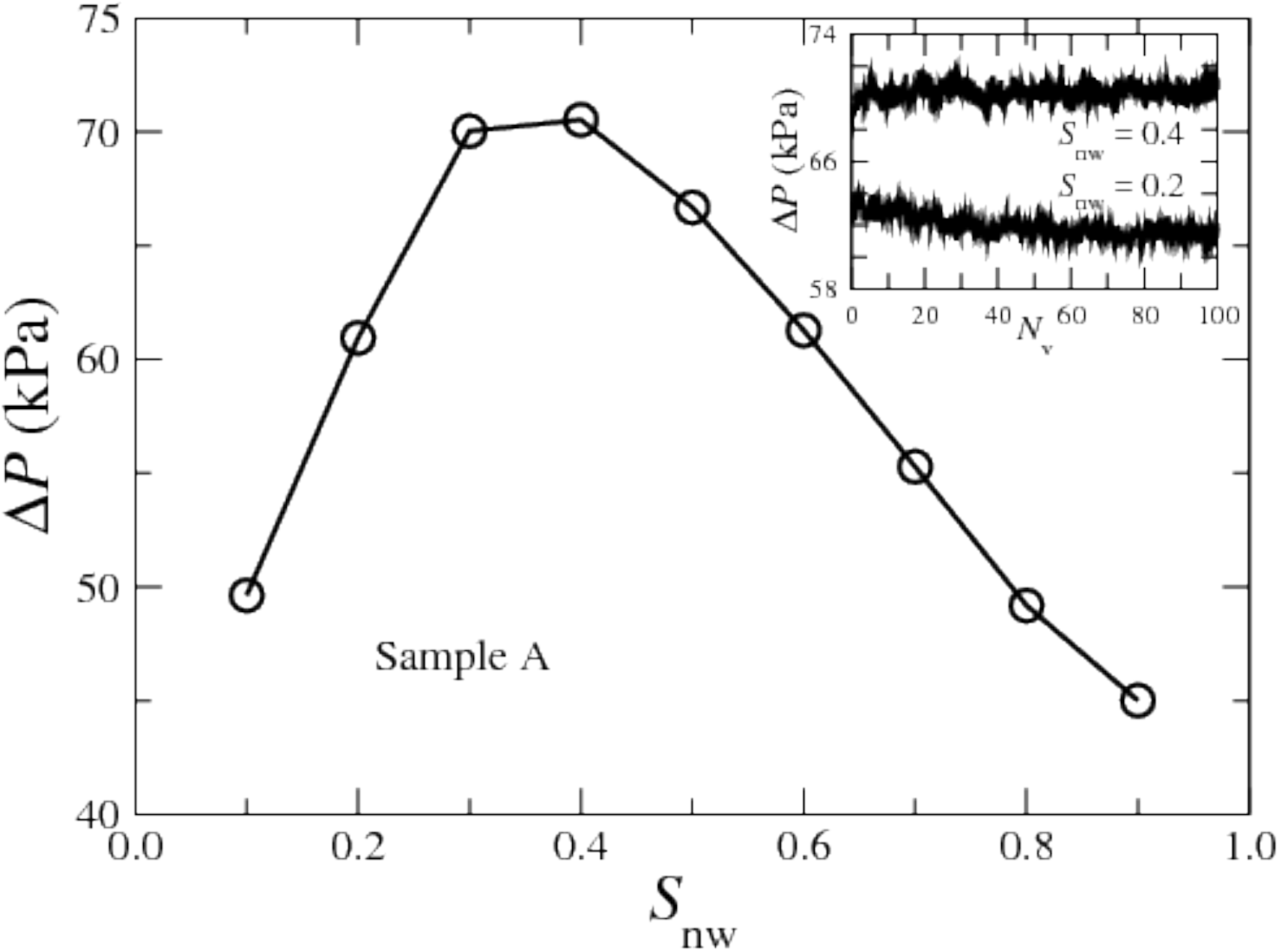}\hfill
    \includegraphics[width=0.32\textwidth,clip]{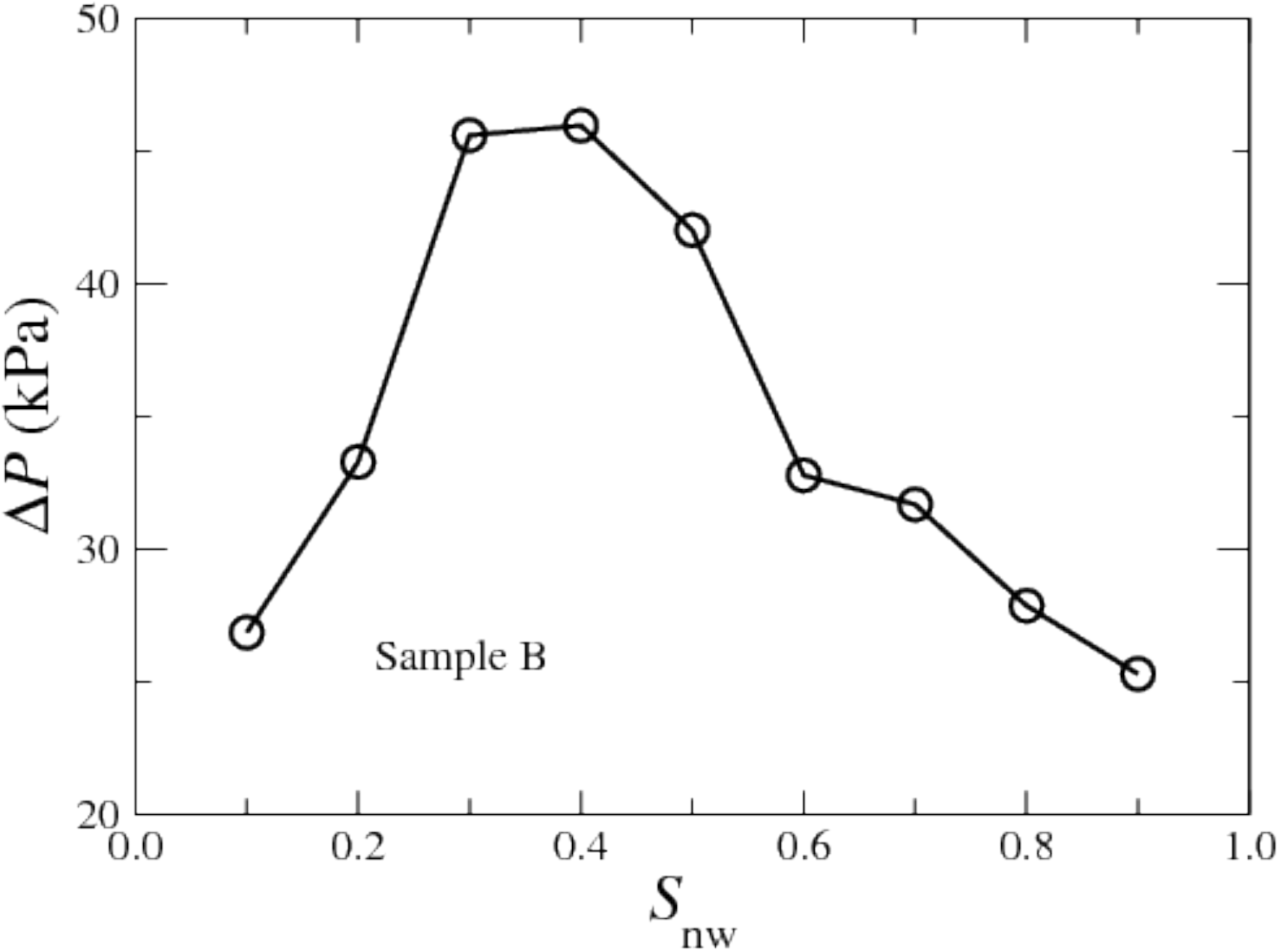}\hfill
    \includegraphics[width=0.32\textwidth,clip]{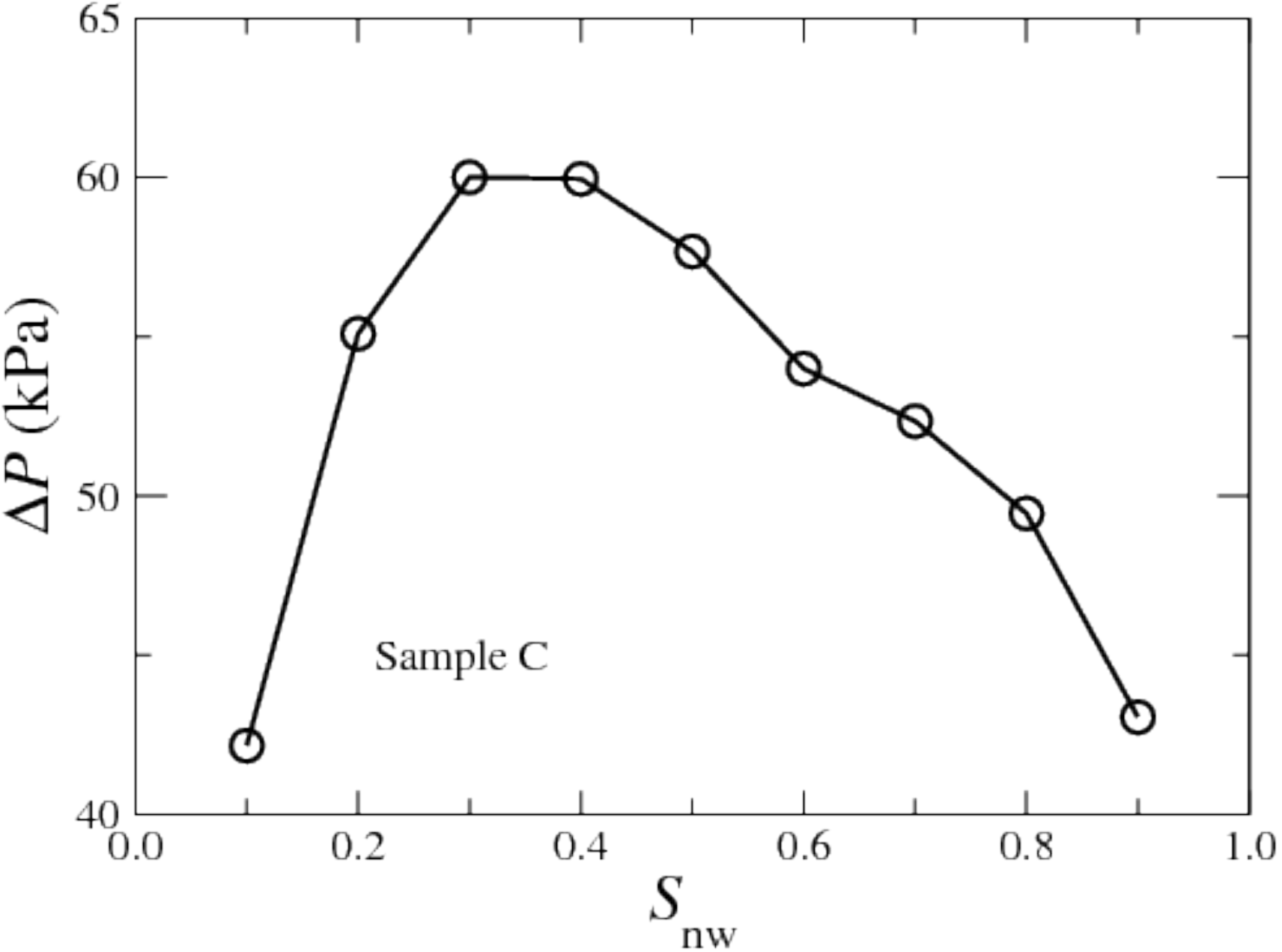}\hfill}
  \caption{\label{ps-steady} Total pressure drop ($\Delta P$) in the
    steady-state for the three networks as a function of
    $S_\text{nw}$. The capillary number $\text{Ca} = 10^{-2}$. $\Delta
    P$ reaches a maximum at an intermediate saturation, which is due
    to the increasing number of interfaces causing higher capillary
    barriers. In the inset of sample A, $\Delta P$ is plotted as a
    function of the pore volumes of fluids passed, which shows the
    evolution of steady state.}
\end{figure}

The two fronts catch up with each other with time and the system
eventually contains a mixture of two fluids in the steady state. In
the steady state, both drainage and imbibition take place at the pore
level and fluid clusters are created, merged and broken up. We
identify the steady state when the average of any measurable
macroscopic quantity stops drifting with time and starts fluctuating
around a constant average value.  Instead of the sequential initial
condition where the two fluids are segregated at two parts of the
system, we may start the simulation from a initial condition where the
two fluids are distributed randomly. In this case the system reaches
to the steady state much faster. Therefore in order to save the
computational time we adopted the random initial condition in all the
following results. When the system evolves into the steady state, we
measure the macroscopic properties with time and take averages of the
measurements.

First we present some fundamental properties of the system by
measuring the non-wetting fractional flow ($F_\text{nw}$) and the
pressure drop across the system ($\Delta P$) as a function of the
saturation. The non-wetting fractional flow is defined as the
proportion of non-wetting fluid flowing across the system given by
$F_\text{nw} = Q_\text{nw}/Q$ where $Q_\text{nw}$ is the volumetric
flow rate of non-wetting fluid through any network cross-section
perpendicular to the overall flow. For a set of miscible fluids with
no capillary forces, $F_\text{nw}$ will be exactly equal to
$S_\text{nw}$ in the steady state for any saturation, which is not the
case here. The measurements of $F_\text{nw}$ in the steady state are
illustrated in Figure \ref{fs-steady}. In the inset of the first plot,
we show the variation of $F_\text{nw}$ as a function of the
pore-volumes of fluids passed through the system which ensures the
steady-state flow situation. The time-averages of $F_\text{nw}$ in the
steady state for the whole range of non-wetting saturation are then
plotted for the three different networks at two capillary numbers
$\text{Ca}=10^{-1}$ and $10^{-2}$. The plots show the well known
S-shape and do not follow the diagonal dashed line corresponding to
$F_\text{nw}=S_\text{nw}$. This is due to the presence of capillary
forces at the interfaces for which the two immiscible fluids do not
flow equally. The one with the higher saturation dominates the flow,
$F_\text{nw}$ is less than $S_\text{nw}$ for low values of
$S_\text{nw}$ and higher than $S_\text{nw}$ for higher values of
$S_\text{nw}$. The curves therefore cross the $F_\text{nw} =
S_\text{nw}$ line at some point, which is not at
$S_\text{nw}=0.5$. This shows the asymmetry between the two fluids, as
one fluid is more wetting than the other with respect to the pore
walls. Moreover, a lower capillary number corresponds to stronger
capillary forces relative to the viscous forces and hence the
deviation of the curves from the diagonal straight line is higher for
$\text{Ca}=10^{-2}$ than for $10^{-1}$ for each network.

\begin{figure}
  \centerline{\hfill
    \includegraphics[width=0.32\textwidth,clip]{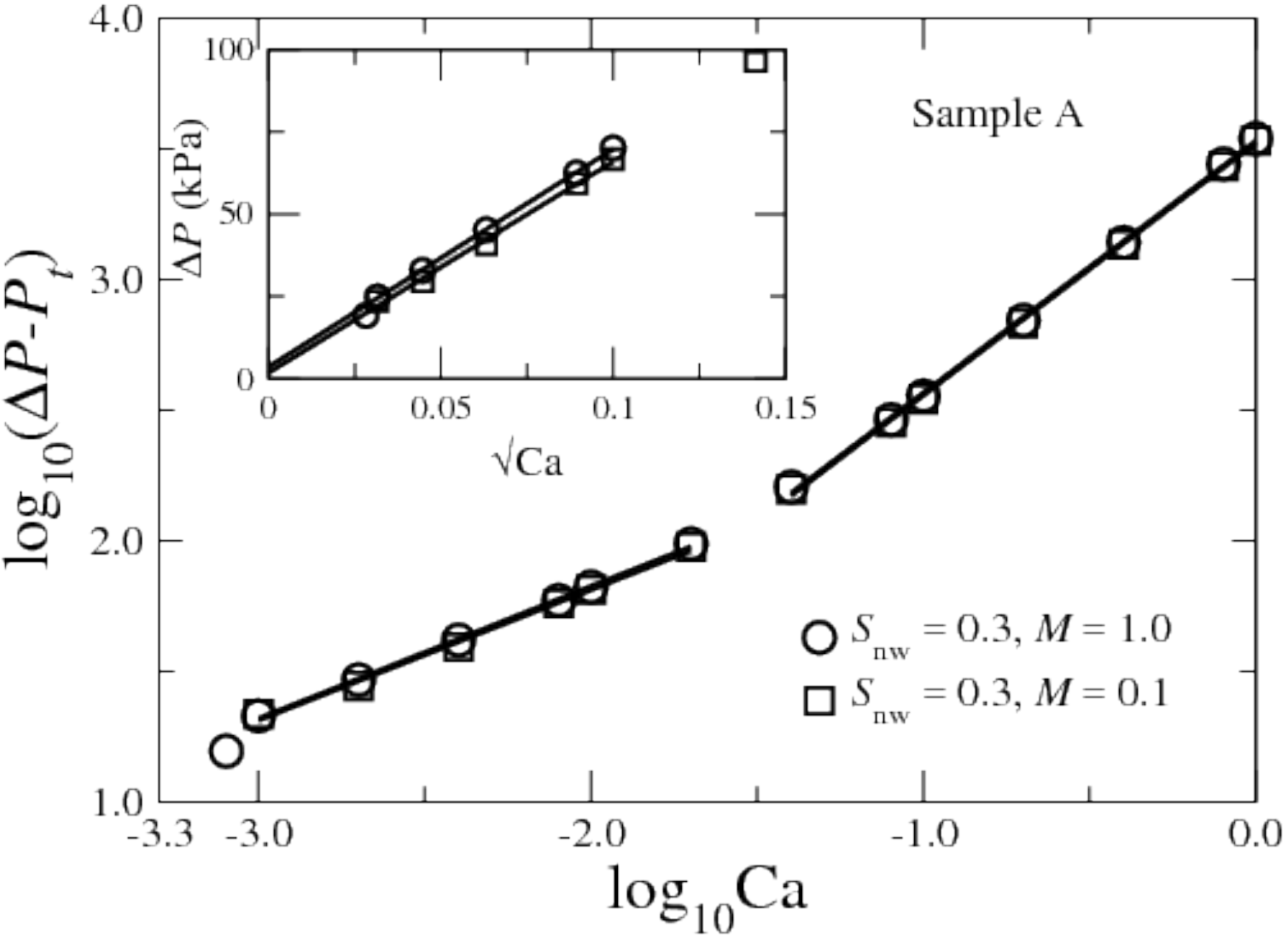}\hfill
    \includegraphics[width=0.32\textwidth,clip]{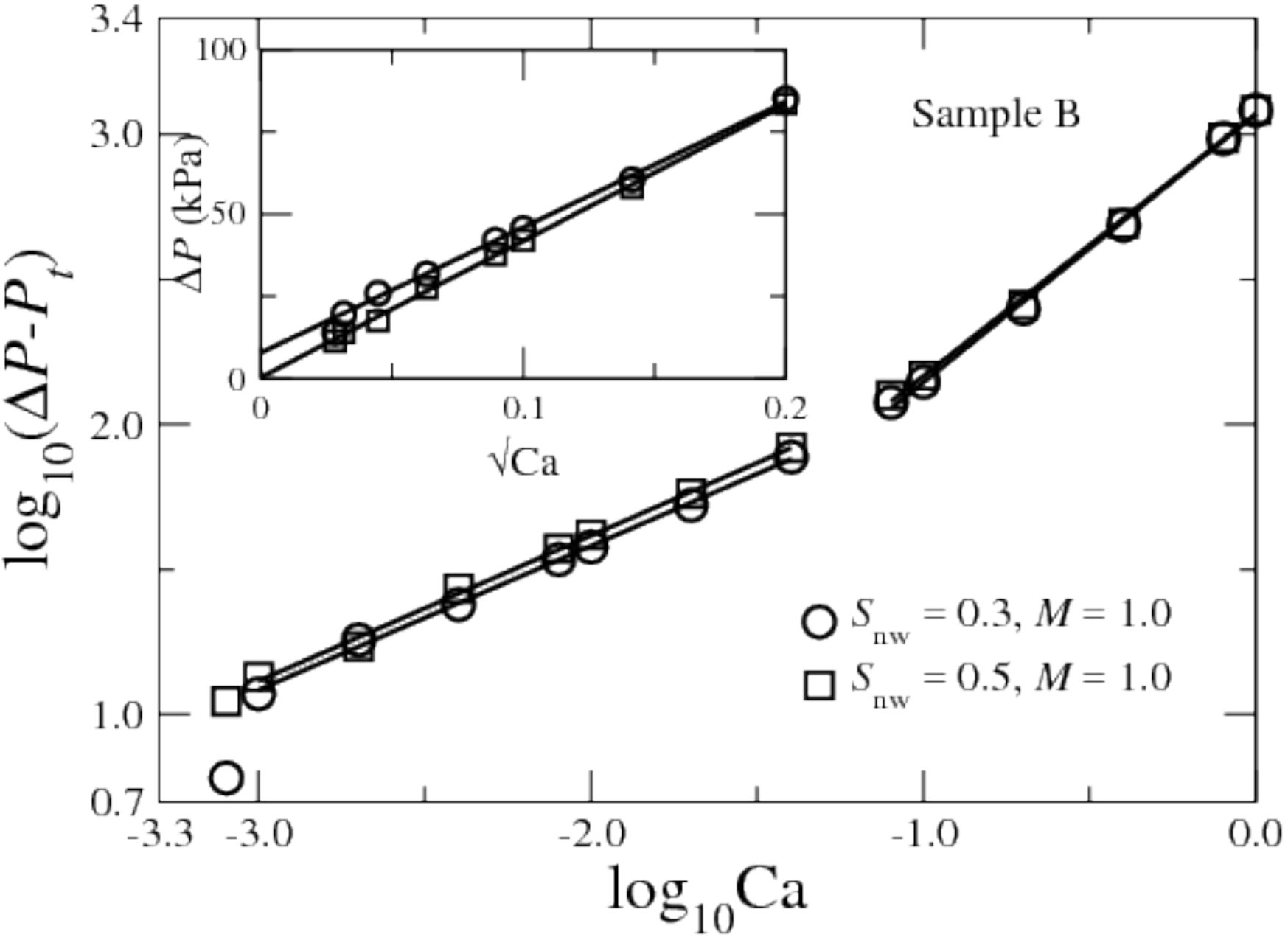}\hfill
    \includegraphics[width=0.32\textwidth,clip]{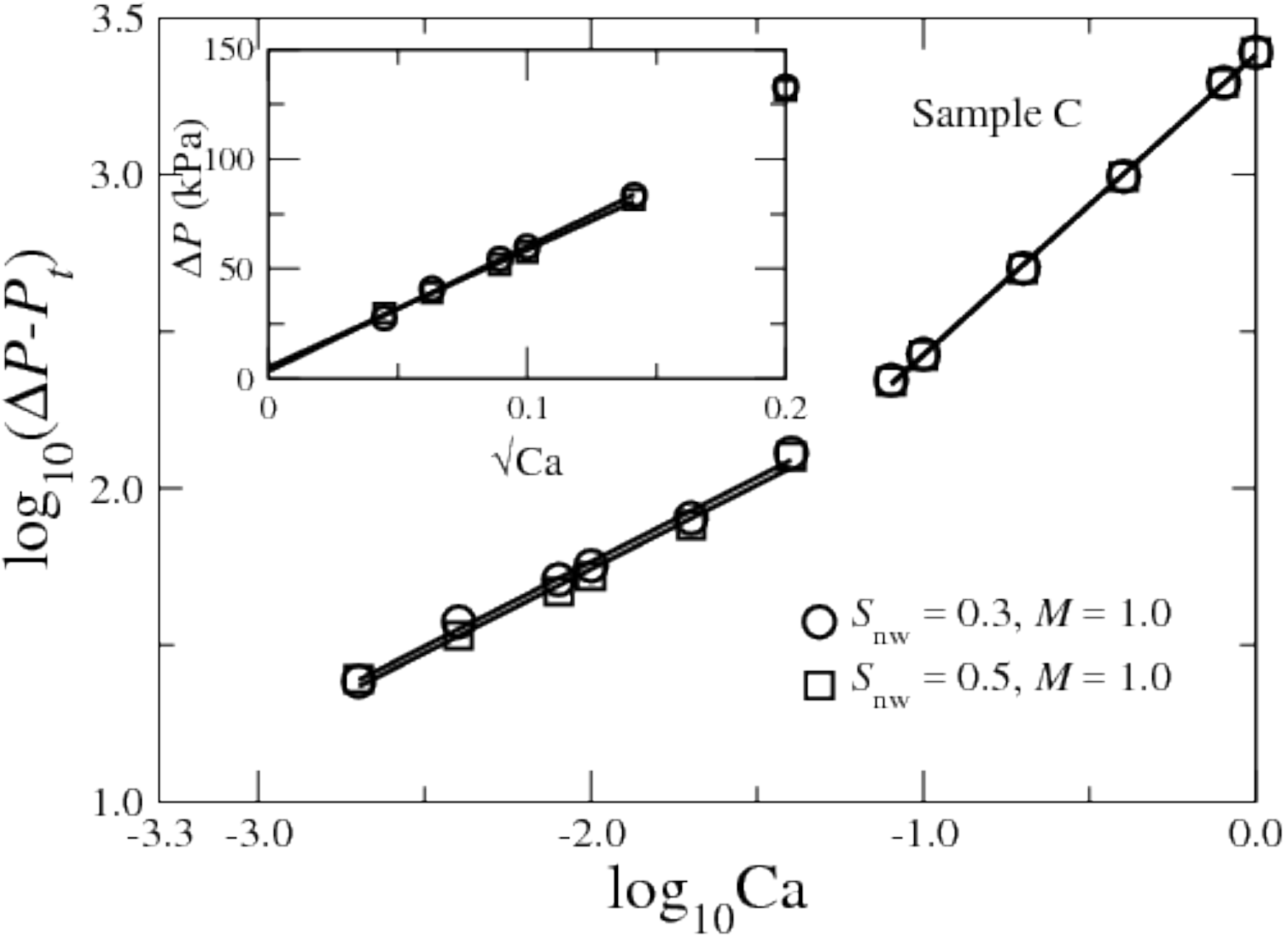}\hfill}
  \caption{\label{pclog} Plot of overall pressure drop as a function
    of the capillary number in the steady state for different
    networks. For network A, results are shown for $S_\text{nw}=0.3$
    with two different viscosity ratios $M=1.0$ and $0.1$. For B and
    C, two different saturations $S_\text{nw}=0.3$ and $0.5$ with
    $M=1.0$ are considered. Measurements of the threshold pressures
    for each simulation are shown in the insets of each figure, where
    $P_t$ is obtained from the $y$-axis intercepts of $\Delta P$ vs
    $\sqrt{\text{Ca}}$ plots. Using the values of $P_t$, the scaling
    exponents are then obtained from the slopes of log-log plots of
    ($\Delta P-P_t$) vs $\text{Ca}$.}
\end{figure}

Variation of the global pressure drop $\Delta P$ as a function of the
saturation $S_\text{nw}$ for constant overall flow rate ($Q$) is shown
in Figure \ref{ps-steady} for the three networks. Here, the capillary
number $\text{Ca}=10^{-2}$ for these plots. The initiation of the
steady state is illustrated in the inset of sample A, where $\Delta P$
fluctuates around an average value in the steady state. The average of
$\Delta P$ in the steady state first increases with the saturation,
reaches a maximum and then decreases again. When $S_\text{nw}$ is
increased from zero, the single phase flow regime, more interfaces
start appearing into the system. This increases the overall capillary
barrier and to keep the same flow rate ($Q$) the pressure drop across
the system needs to be increased. $\Delta P$ is maximum at some
intermediate saturation and then starts decreasing again as
$S_\text{nw}$ approaches to $1$, as the system again approaches to the
single phase flow regime. Interestingly, the maximum of $\Delta P$ is
not at $S_\text{nw}=0.5$, but close to the saturation where
$F_\text{nw}$ crosses the diagonal $F_\text{nw}=S_\text{nw}$ line as
observed in Figure \ref{fs-steady}. This was predicted in
\cite{hsbksv16}.

\begin{table}
  \centering
  \begin{tabular}{c|c|c|c|c|c}
    \hline
    Sample            & $M$ & $S_\text{nw}$ & $P_t$ (kPa) & $\alpha$ (Low $\text{Ca}$)& $\alpha$ (High $\text{Ca}$) \\
    \hline
    \multirow{2}{*}{A}& $1.0$  & $0.3$ & $3.543$   & $0.51\pm 0.01$ & $0.96\pm 0.01$ \\
                      & $0.1$  & $0.3$ & $1.745$   & $0.50\pm 0.02$ & $0.97\pm 0.01$ \\
    \hline
    \multirow{2}{*}{B}& $1.0$  & $0.3$ & $7.795$   & $0.50\pm 0.01$ & $0.92\pm 0.02$ \\
                      & $1.0$  & $0.5$ & $0.343$   & $0.50\pm 0.01$ & $0.90\pm 0.02$ \\
    \hline
    \multirow{2}{*}{C}& $1.0$  & $0.3$ & $3.354$   & $0.54\pm 0.02$ & $0.96\pm 0.01$ \\
                      & $1.0$  & $0.5$ & $5.183$   & $0.54\pm 0.03$ & $0.96\pm 0.01$ \\
    \hline
  \end{tabular}
  \caption{\label{pcexpo} Values the threshold pressures $P_t$ and the
    corresponding scaling exponents $\alpha$ for different
    simulations. The threshold pressures are obtained from the
    intercepts in the $y$-axis of $\Delta P$ vs $\sqrt{\text{Ca}}$
    plots as shown in the insets of Figure \ref{pclog}. Using the
    values of $P_t$, the scaling exponents are then obtained from the
    slopes of $\log(\Delta P-P_t)$ vs $\log\text{Ca}$ plots.}
\end{table}

We now present the simulation results related to the scaling of
$\Delta P$ with $Q$ in the steady-state. The values of $\text{Ca}$
range from $\approx 10^{-3}$ to $1$ in our simulations. We observed
that the crossover from the non-linear to the linear scaling falls in
this range and we can see both the scaling regimes. We therefore have
not performed simulations around $\text{Ca}=10^{-5}$ as we did in our
experiments.  Results are illustrated in Figure \ref{pclog} where we
plot $\log(\Delta P-P_t)$ as a function of $\log\text{Ca}$. First, we
need to find the threshold pressure $P_t$ for each case. In our
experiments, $P_t$ was found by measuring the differential pressure at
$Q=0$. In the simulations for the Bingham fluid \cite{roux87}, $P_t$
was determined by decreasing the external current (or flow-rate) from
a large value and identifying the flow-paths with a search
algorithm. This procedure is not feasible for dynamic two-phase flow
network models, as the interfaces move with time and consequently the
flow-paths change. In the case of the two-phase flow simulations in a
2D regular network \cite{sinhaepl12}, $P_t$ was measured by minimizing
the linear least-square fit errors of $\log(\Delta P-P_t)$ versus
$\log\text{Ca}$ data. There, the numerical results were averaged over
different samples and time, whereas in the case of the reconstructed
3D network we only have one network per sample and do not have
opportunity to do sample averaging. This leads to higher statistical
fluctuations in the error measurement and therefore finding $P_t$
based on the minimum error was not possible. Therefore, as $P_t$ is
the value of $\Delta P$ as $Q\to 0$, we calculated $P_t$ directly from
the numerical data of $\Delta P$ versus $Q$. For the low $\text{Ca}$
regime, we expect $Q\sim(\Delta P-P_t)^2$ and therefore we plot
$\Delta P$ as a function of $\sqrt \text{Ca}$ which leads to straight
lines for the low $\text{Ca}$ regime for each sample. The plots are
shown in the insets of each plot in Figure \ref{pclog}. Values of the
intercepts of the straight lines in the $y$-axis corresponds the
threshold pressure $P_t$. Using these values of $P_t$ we plot
$\log(\Delta P-P_t)$ as a function of $\log\text{Ca}$ in Figure
\ref{pclog} for the three samples for different values of
$S_\text{nw}$ and different viscosity ratios. From the slopes, we see
two distinct regimes of flow with two different slopes and there is a
sharp crossover in between the two regimes. The results are summarized
in in Table \ref{pcexpo}. For the low $\text{Ca}$ regime all the
slopes are close to $0.5$ which lead to $Q\sim (\Delta P-P_t)^2$ as
shown in equation \ref{effdarcy}. For the high flow rates, all the
slopes are close to $1$ and the flow is Newtonian.

An interesting aspect we observe concerns the transition point from
the nonlinear low-$\text{Ca}$ regime to the linear high-$\text{Ca}$
regime. This transition point seems to vary between studies. In the
experiments, we find it around $\text{Ca}\approx 10^{-4.75}$ while our
simulations show it around $\text{Ca}\approx 10^{-1.5}$. Even in the
simulations, the transition point seems to vary slightly among
different porous media samples A, B, and C. Furthermore, Tallakstad
{\it et al.} did not observe the transition for $\text{Ca}<10^{-1}$
and therefore it should be at a more higher value
\cite{tallakstad09}. It seems that the transition point is strongly
determined by geometric and physical characteristics of the porous
medium itself. At this point, we cannot answer how this transition
point depends on the network characteristics, saturation or viscosity
ratio and will be an interesting aspect to explore in the future.

\section{Conclusions}
In this article, we presented our experimental and numerical study to
investigate the relationship between the pressure drop and the
volumetric flow rate in the steady-state two-phase flow of immiscible
fluids in three dimensional porous media. Our two-phase flow
experiments utilize a three-dimensional porous medium made of glass
beads with air and de-ionized water flowing through it. We performed
numerical simulations constructing a network model of two-phase flow
in 3D reconstructed pore networks. We have addressed the different
nonlinear relationships that were observed in 2D \cite{tallakstad09}
and 3D \cite{rassi11} experiments and in 2D simulations
\cite{sinhaepl12} reported earlier. With our experiments and
simulations, we show that the capillary pressures at the interfaces in
between the immiscible fluids introduce a pressure barrier at each
pore and the presence of these barriers in the disorder pore-network
effectively creates a yield threshold in the system, making the fluids
to behave like a Bingham viscoplastic fluid in a network
\cite{roux87,talon14,talon15}. There are two regimes of the flow, in
the regime where the capillary forces are comparable with the viscous
forces, the flow rate varies quadratically with the excess pressure
drop. With the increase in the flow-rate, the capillary pressures
become negligible and there is a crossover into a linear flow
regime. Both these two flow regimes are well demonstrated by our
experiments and simulations in 3D. Our results here in 3D are also in
the agreement with the mean-field theory presented earlier
\cite{sinhaepl12} which shows that the quadratic scaling does not
depend on the dimensionality of the pore network.

We have considered $S_\text{nw}=0.3$ and $0.5$ in the simulations and
$F_\text{nw}=0.5$ in the experiments which are in the intermediate
range of saturation or fractional flow. In this regime, both the
fluids contribute to the flow and a large number of interfaces exist
in the system introducing capillary barriers at each pore. If one
moves towards $S_\text{nw}\to 0$ or $1$, the system will drift to the
single-phase flow regime and eventually the interfaces as well as the
capillary barriers will disappear. This should drive the system
completely into the linear Newtonian flow regime. The quadratic flow
regime we have seen here should therefore disappear as $S_\text{nw}\to
0$ or $1$ and the linear flow regime will cover the whole range of
$\text{Ca}$. It will be interesting to study this transition as a
function of the saturation or the fractional flow by exploring the
whole parameter space. However this needs an enormous range of
experiments and simulations as approaching steady state needs many
pore volumes of the fluid to be reached.  Recently, a Monte Carlo (MC)
algorithm has been proposed for the network models of two-phase flow
in 2D networks \cite{savanimc} and we look forward towards further
development of the MC algorithm for the 3D networks in order to have
an efficient way to study the effective scaling of the pressure and
the flow rate of steady-state two-phase flow for the whole parameter
space in the future.

\acknowledgement{We thank Eirik Grude Flekk{\o}y, Knut J{\o}rgen
  M{\aa}l{\o}y, Laurent Talon, Signe Kjelstrup, Dick Bedeaux and
  Morten Vassvik for many interesting discussions. C. Prather,
  J. Bray, L. Thrane and S. Codd acknowledges support by the National
  Science Foundation under CBET grant 1335534. A. Bender and
  M. Danczyk acknowledge support from the Undergraduate Scholars
  Program at Montana State University. K. Keepseagle acknowledges
  support from the McNairs Program and from Grant Number P20 RR-16455
  from the National Center for Research Resources (NCRR), a component
  of the National Institutes of Health (NIH). A. Hansen thanks the
  Beijing Computational Science Research Center and its director,
  Professor Hai-Qing Lin, for financial support and for providing an
  excellent atmosphere for doing science.}


\end{document}